\documentclass[a4paper, traditabstract]{aa} 
\usepackage{graphicx}
\usepackage{txfonts}
\usepackage{url}
\usepackage{natbib}
\usepackage{amssymb}
\usepackage{booktabs}
\usepackage{subfig}
\usepackage{commath}
\usepackage[usenames,dvipsnames]{xcolor}
\usepackage[bookmarks, colorlinks, breaklinks, 
]{hyperref}  
\hypersetup{linkcolor=blue,citecolor=MidnightBlue,filecolor=black,urlcolor=MidnightBlue} 

\usepackage{microtype}
\usepackage{breqn}
\def\msol{{M$_{\odot}$}}

\def\xmm{XMM-{\it Newton}}

\def\LX {L_{500}}
\def\Lv {L_{500}}
\def\RV {R_{500}}
\def\MV {M_{500}}
\def\YX {Y_{\rm X}}
\def\TX {T_{\rm X}}

\def \xij {x_{\rm i,j}}

\def \sigdij {\sigma^2_{\rm i,j}}

\newfont{\gwpfont}{cmssq8 scaled 1000}
\newcommand{\rexcess}{{\gwpfont REXCESS}}
\newcommand{\excpres}{{\gwpfont EXCPReS}}
\def \chandra {\hbox{\it Chandra}}

\begin{document}
 
\title{Evolution of X-ray galaxy Cluster Properties in a Representative Sample (EXCPReS)}
\subtitle{Optimal binning for temperature profile extraction}
\author{C.M.H. Chen \inst{1}, 
M. Arnaud \inst{1}, 
E. Pointecouteau \inst{2}, 
G.W. Pratt \inst{1}, and 
A. Iqbal  \inst{1}}
\authorrunning{C.M.H. Chen et al.}
\titlerunning{Evolution of X-ray galaxy Cluster Properties in a Representative Sample (EXCPReS)}
 \institute{
 $^1$ Université Paris-Saclay, Université Paris Cité, CEA, CNRS, AIM, 91191, Gif-sur-Yvette, France \\
 \email{gabriel.pratt@cea.fr} \\ 
 $^2$ IRAP, Université de Toulouse, CNRS, CNES, UPS, 9 Av. colonel Roche, BP 44346, F-31028 Toulouse Cedex 4, France\\
 \email{etienne.pointecouteau@irap.omp.eu} \\ 
}
\date{Received ; accepted }
\abstract{
We present \xmm\ observations of a representative X-ray selected sample of 31 galaxy clusters at moderate redshift  $(0.4<z<0.6)$, spanning the mass range $10^{14} < M_{\textrm 500} < 10^{15}$~M$_\odot$. This sample, \excpres\ (Evolution of X-ray galaxy Cluster Properties in a Representative
Sample), is used to test and validate a new method to produce optimally-binned cluster X-ray temperature profiles. The method uses a dynamic
programming algorithm, based on partitioning of the soft-band X-ray surface brightness profile, to obtain a binning scheme that optimally fulfils a given signal-to-noise threshold criterion out to large radius.  
From the resulting optimally-binned \excpres\ temperature profiles, and combining with those from the local \rexcess\ sample, we provide a generic scaling relation between the relative error on the temperature and the [0.3-2]\, keV surface brightness signal-to-noise ratio, and its dependence on temperature and redshift. We derive an average scaled 3D temperature profile for the sample. Comparing to the average scaled 3D temperature profiles from \rexcess, we find no evidence for evolution of the average profile shape within the redshift range that we probe.}


   \maketitle


\section{Introduction}
Galaxy clusters form through the gravitational collapse of the dominant dark matter component, with the gas of the intra-cluster medium (ICM) ‘following’ the gravitational potential as the object grows by accretion and merging. The ICM is heated to X-ray emitting temperatures by shocks and compression during this hierarchical assembly process under gravity. Feedback from active galactic nuclei, supernovae, and gas cooling further modify the properties of the ICM over cosmic time \citep{kra12,vog14, sch15, sch23}.

In this context, spatially resolved measurements of the thermodynamic properties of the ICM contain crucial information on the physics governing the formation and evolution of groups and clusters. Radial profiles of gas density, $n_{\rm e}$, and temperature, $kT$, have become fundamental tools for measurement of key quantities such as pressure, $P$, entropy, $K$, and hydrostatic mass, and for making comparisons with predictions from numerical simulations \citep[see, e.g.][for recent reviews]{lov22,kay22}. However, while the gas density is simple to obtain from X-ray imaging, determination of a radial temperature profile requires an annular binning scheme that yields sufficient signal to build and model the spectrum. The problem is compounded by the density-squared dependence of the X-ray emission, a typical cluster emissivity profile that steepens dramatically with radius, and by the drop in the signal-to-noise (S/N) at lower masses and higher redshifts.

The radial temperature profiles of local ($z \lesssim 0.3$) clusters and groups are now well characterised \citep[][]{mar98,deg02,vik05,vik06,pra07,bal07,lec08,sun09,arn10,lov15}. The temperature declines gradually towards the outer regions from a peak at $R \sim 0.2\,R_{500}$. In the inner regions, non-cool core systems are typically approximately isothermal at the peak temperature, while cool core systems exhibit a characteristic smooth drop to $1/2 - 1/3$ of the peak temperature value. In particular, \citet{vik06} and \citet{sun09} showed the remarkable similarity and tight scaling with mass in relaxed systems. The Representative \xmm\ Cluster Structure Survey (\rexcess), a representative sample of X-ray selected clusters, has provided the mean pressure profile, its dispersion, and the mass scaling \citep{arn10} that is used in all matched
multi-filter Sunyaev-Zeldovich effect (SZE) survey detection algorithms \citep[e.g.][]{mel06}. It has also yielded strong constraints on the radial and mass dependence of the entropy distribution \citep{pra10}. 

Until recently however, individual radial temperature profiles were rarely available beyond $z > 0.3$ (however, see e.g. \citealt{kot05,bal12,man16} for pioneering studies). While the advent of SZE surveys by ACT \citep{has13}, SPT \citep{ble15} and Planck \citep{PSZ2}, has transformed the quest for high-$z$ systems,
X-ray follow-up deep enough to measure annular temperature profiles of these new $z>0.3$ SZE-discovered clusters has concentrated on the highest-mass objects. Being X-ray bright, they are the `easiest' systems to observe, leading to good a precision on the radial temperature distribution \citep[see e.g.][where the individual thermodynamic profiles of systems up to $z \sim 1$ are measured]{bar17,bar19}. Results from stacking \citep[e.g.][]{mac14} can yield the average behaviour, but do not offer insights into the intrinsic scatter in the profiles. 

In this paper we introduce the Evolution of X-ray galaxy Cluster Properties in a Representative Sample (\excpres). Consisting of 31 X-ray-selected clusters in the redshift range $0.4 < z < 0.6$ and the mass range $0.1-1.3 \times 10^{15}$ M$_\odot$, \excpres\ was designed to be a moderate-redshift analogue of \rexcess. We use \excpres\ to test and validate a novel method to optimally bin X-ray data to reconstruct the 3D temperature distribution of the ICM. 
In the following, we present the \excpres\ sample, the \xmm\ observations and data analysis. We describe our method for binning of the surface brightness profiles to derive optimal temperature profiles. 
 The method is applied to the \excpres\ sample, and the resulting profiles, and their dispersion, are compared to those of the local \rexcess\ sample.
Throughout this paper we assume a flat $\Lambda$CDM model with $H_0=70$~km~s$^{-1}$~Mpc$^{-1}$, $\Omega_m=0.3$ and $\Omega_\Lambda=0.7$. Uncertainties are quoted at the $68\%$ confidence level. The variables $\MV$ and $\RV$ are the total mass within $\RV$ and radius corresponding to a total density contrast of $\Delta = 500 \, \rho_{\rm c}(z)$, where $\rho_{\rm c}(z)$ is the critical density of the universe at the cluster redshift.

\section{The sample}
\label{s:samp} %

\begin{figure}[!t]
\includegraphics[width=0.9\columnwidth]{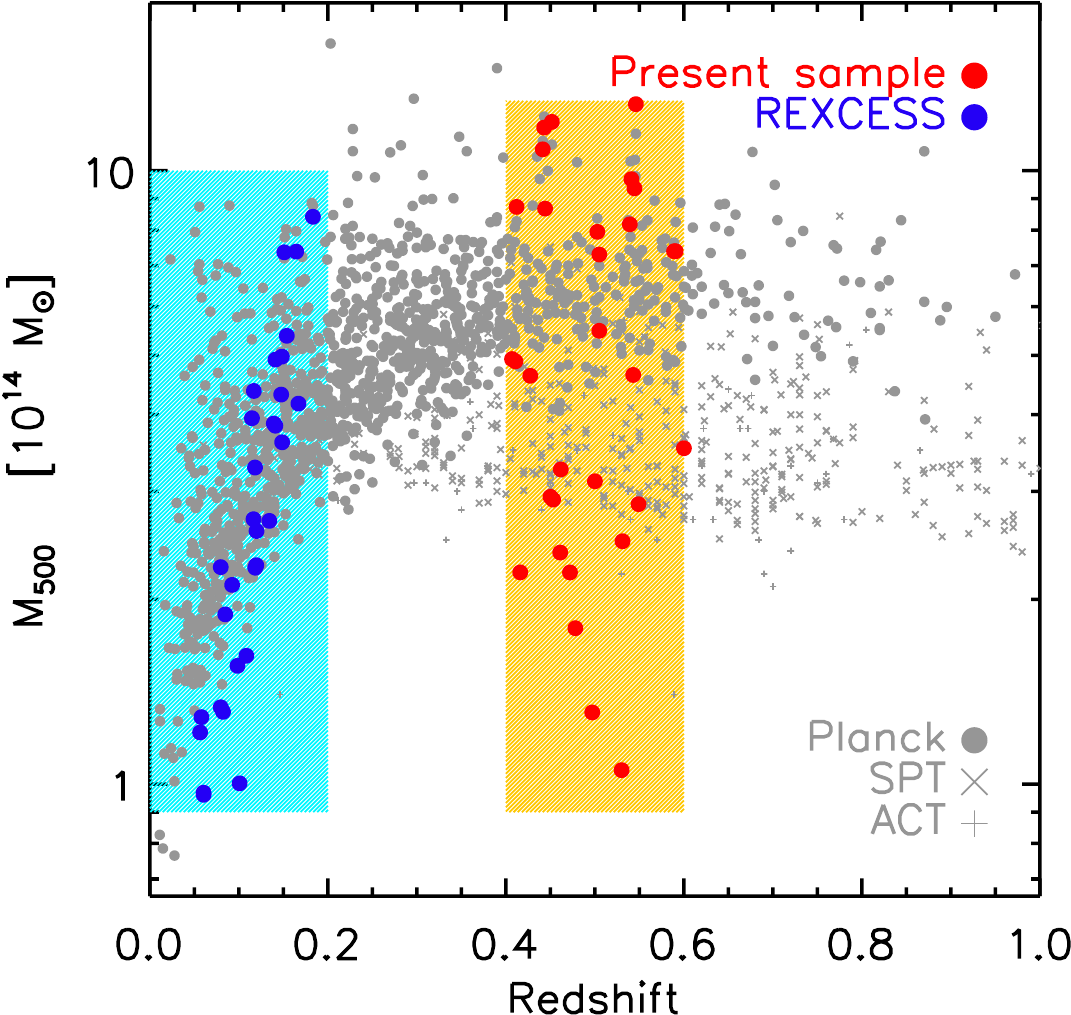}
\caption{\footnotesize Distribution in the $M-z$ plane of the \excpres\ sample (red points). 
The local X-ray-selected \rexcess\ sample \citep{boe07} is shown with blue points. Shown for comparison are confirmed clusters from major SZE surveys from which individual spatially resolved temperature profiles are measurable: Filled circles: {\it Planck}  clusters  \citep{esz,psz1,PSZ2}. Crosses: SPT \citep{ble15}. Plus symbols: ACT \citep{has13}. }
\label{f:Mz}
\end{figure}

Our aim is to obtain spatially-resolved temperature measurements at moderate redshift across the full cluster mass range (i.e. $M_{500} \gtrsim 10^{14}$ M$_{\odot}$). As SZE-selected clusters typically probe higher masses (see Fig.~\ref{f:Mz}), we chose to focus instead on X-ray selected clusters at a median redshift of $z = 0.5$. A logical local reference in this case is the \rexcess\ sample \citep{boe07}, which covers a similar mass range, but at lower redshifts ($0.055 < z < 0.183$). The chosen
median redshift of \excpres\ is the highest $z$ for which we can define a complete
sample covering the whole cluster mass range of the various ROSAT surveys. Indeed, clusters below $\sim 10^{14}$ M$_{\odot}$ have such low luminosities that at $z\gtrsim
0.6$ they begin to fall rapidly below the detection limits of these surveys.

\subsection{Parent MCXC sample}
Our sample is drawn from the Meta Catalogue of X-ray detected Clusters of galaxies \citep[MCXC;][]{pif11}. MCXC is based on the Einstein Medium Sensitivity Survey \citep[EMSS][]{gio90,hen04} and on the ROSAT All-Sky and Serendipitous surveys. We used the MCXC-II, which includes updated redshifts and $\sim500$ additional clusters \citep{sad24}. Redshift revision was based on comparison of the redshifts from the NED and Simbad data bases, and cross-matching with cluster catalogues extracted from large optical surveys, in particular the SDSS \citep[e.g.][]{wen15,ryk16}.

\begin{figure}[!t]
\includegraphics[width=0.9\columnwidth]{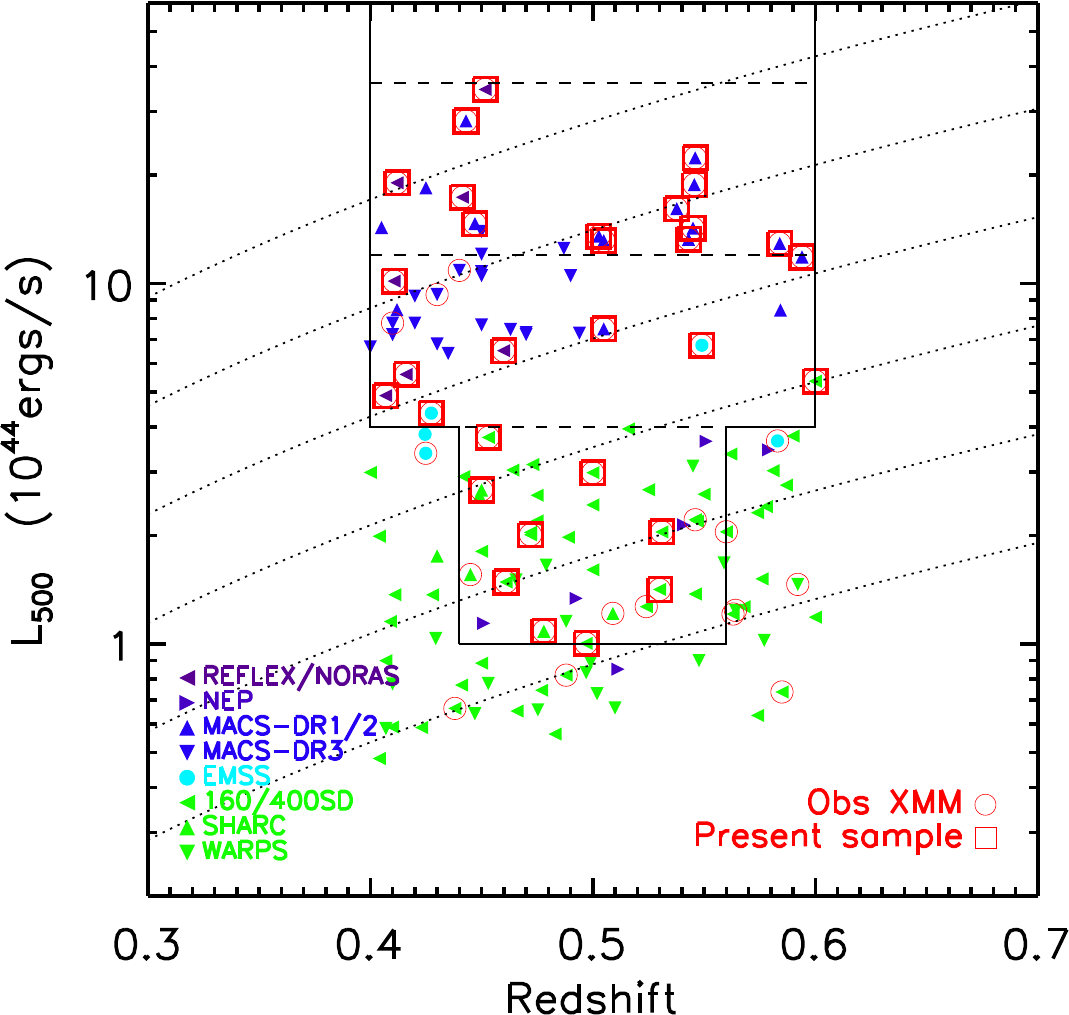}
\caption{\footnotesize Distribution of the MCXC clusters in the $z$--$\LX$ plane \citep{pif11}. Redshift and $\LX$, the luminosity within $\RV$,  are  from the updated MCXC catalogue \citep{sad23}. Each cluster is colour-coded according to the  survey from which the luminosity is taken. Clusters with \xmm\ observations are marked with red circles and those included in the \excpres\ sample are marked with red boxes. The dotted lines mark the [0.1-2.4]~keV band flux taking into account the $K_z$ correction for a typical gas temperature of 5~keV. Levels are separated by 2~dex, from $1.25\times 10^{-13}$ up to  $4\times 10^{-12}$~ergs~s$^{-1}$~cm$^-2$.}
\label{f:lxz}
\end{figure}

\begin{table*}[!t]
\caption{\footnotesize \excpres\ sample and \xmm\ observations. Columns are 1: object index. 2-5: Cluster MCXC name, other name, detection survey and redshift. 6-7: Right ascension and declination in 2000 equinox of X--ray peak in \xmm\  image. 6: \xmm\ OBSID. 7: EPIC (total EMOS and EPN) effective exposure time (i.e. after flare cleaning). }
\begin{center}
\begin{minipage}{\textwidth}
\resizebox{\textwidth}{!} {
\begin{tabular}{llllcccccc}
\toprule
ID & Name & Other Name & Catalogue & z & RA & DEC & OBSID & Exp. \\
&& & & & (h,m,s) & (d,am,as) & & MOS/PN  \\
 && & & &  &  & & ksec  \\
\midrule
  1& MCXC J0018.5+1626 & MACS J0018.5+1626; CL 0016+1609 & MACS; EMSS&0.541 &  0 18 33  &+16  26  08  &	0111000101 &  33.3/23.9 \\
  & &  & & &   &  &	0111000201 &   \\
   2&MCXC J0030.5+2618 & [BVH2007] 2; [VMF98] 1 & 400SD; WARPSII; 160SD&    0.500 &  0 30 33  &+26  18  06  &	0402750201 &  26.9/ 19.7 \\
   3&MCXC J0221.1+1958 & RX J0221.1+1958& SHARC BRIGHT&    0.450 &  2 21 09  &+19  58  03  &	0402750301 &  25.8/ 16.0 \\
   4&MCXC J0257.1-2326 & MACS J0257.1-2325& MACS&    0.505 &  2 57 08  &-23  26  07  &	0551850201 &  25.6/  9.9 \\
   5&MCXC J0417.5-1154 & MACS J0417.5-1154& MACS&    0.443 &  4 17 34  &-11  54  32  &	0827310101 &   9.9/  4.7 \\
   6&MCXC J0454.1-0300 & MACS J0454.1-0300;  MS 0451.6+0305  & MACS; EMSS&    0.539 &  4 54 10  &-03  00  57  &	0205670101 &  22.0/17.2 \\
   7&MCXC J0522.2-3624 &[VMF98] 41 & 400SD ;160SD&    0.472 &  5 22 15  &-36  25  11  &	0302580901 &  18.2/14.4 \\
   8&MCXC J0647.8+7014 & MACS J0647.7+7015	& MACS&    0.591 &  6 47 50  &+70  14  53  &	0551850401 &  71.4/44.6 \\
   & & & & & & & 0551851301 & \\
   9&MCXC J0717.5+3745 & MACS J0717.5+3745& MACS&    0.546 &  7 17 31  &+37  45  31  &	0672420101 & 156.5/116.5 \\
   & & & & & & & 0672420201 & \\
   & & & & & & & 0672420301 & \\
   10&MCXC J0856.1+3756 & RXC J0856.1+3756& NORAS&0.411 &  8 56 12  &+37  56  09  &	0302581801 &  18.3/ 6.6 \\
   11&MCXC J0911.1+1746 & MACS J0911.2+1746& MACS&0.505 &  9 11 11  &+17  46  33  &	0693662501 &  33.0/25.5 \\
   12&MCXC J0943.1+4659 & RXC J0943.1+4659; A0851  & NORAS&    0.407 &  9 43 00  &+46  59  28  &	0106460101 &  32.4/20.7 \\
   13&MCXC J0957.8+6534 & RXC J0957.8+6534& 160SD&    0.530 &  9 57 51  &+65  34  24  &	0502430201 &  37.4/22.0 \\
   14&MCXC J1003.0+3254 & RXC J1003.0+3254& NORAS; 400SD&    0.416 & 10 03 04  &+32  53  44  &	0302581601 &  22.2/ 7.2 \\
   15&MCXC J1120.1+4318 & RX J1120.1+4318& 400SD; SHARC BRIGHT; WARPSII & 0.600 & 11 20 07  &+43  18  07  &	0107860201 &  17.7/10.7 \\
   16&MCXC J1149.5+2224 & MACS J1149.5+2223& MACS; NORAS 				&    0.544 & 11 49 35  &+22  24  03  &	0693661701 &  12.5/ 2.1 \\
   17&MCXC J1206.2-0848 & RXC J1206.2-0848         & REFLEX; MACS   			&    0.441 & 12 06 12  &-08  48  00  &	0502430401 &  29.1/21.0 \\
   18&MCXC J1244.0+1653 & MS1241.5+1710 			& EMSS 						&    0.549 & 12 44 01  &+16  53  43  &	0302581501 &  28.7/19.0 \\
   19&MCXC J1311.5-0551 & RX J1311.5-0551     		& 160SD						&    0.461 & 13 11 30  &-05  52  03  &	0302582201 &  21.4/17.0 \\
   20&MCXC J1347.5-1144 & RX J1347.5-1145  		& REFLEX; MACS   			&    0.452 & 13 47 30  &-11  45  08  &	0112960101 &  32.3/25.4 \\
   21&MCXC J1411.4+5212 & RXC J1411.4+5212	3C295  	& NORAS						&    0.462 & 14 11 20  &+52  12  10  &	0804271501 &  15.3/12.2 \\
   22&MCXC J1423.7+2404 & MACS J1423.8+2404   		& MACS						&    0.543 & 14 23 48  &+24  04  43  &	0720700401 &  23.1/19.4 \\
   23&MCXC J1623.5+2634 & MS1621.5+2640			& EMSS  				&    0.427 & 16 23 35  &+26  34  20  &	0112190701 &   4.3/0.0 \\
   24&MCXC J1701.3+6414 &	RXC J1701.3+6414						& 400SD; SHARC BRIGHT; 160SD &    0.453 & 17 01 23 &+64  14  09  &	0723700201 &  35.3/30.5 \\
   25&MCXC J2129.4-0741 & MACS J2129.4-0741& MACS&0.589 & 21 29 26  &-07  41  28  &	0700182001 &   8.6/3.3 \\
   26&MCXC J2146.0+0423 & RX J2146.0+0423& 160SD; WARPS& 0.531 & 21 46 05  &+04  23  00  &	0302580701 &  20.7/17.2 \\
   27&MCXC J2214.9-1400 & MACS J2214.9-1359&MACS&    0.503 & 22 14 57  &-14  00  16  &	0693661901 &17.4/ 7.1 \\
   28&MCXC J2228.6+2036 & RXC J2228.6+2036& NORAS; eBCS&    0.412 & 22 28 33  &+20  37  13  &	0147890101 &  23.2/14.6 \\
   29&MCXC J2243.3-0935 & MACS J2243.3-0935& MACS&    0.444 & 22 43 21  &-09  35  40  &	0503490201 & 102.9/77.2 \\
   30&MCXC J2328.8+1453 & RX J2328.8+1453& 160SD; WARPSII&    0.497 & 23 28 51  &+14  53  01  &	0502430301 &  85.7/62.9 \\
   31&MCXC J2359.5-3211 & RX J2359.5-3211& SHARC SOUTH&    0.478 & 23 59 36  &-32  11  17  &	0302580501 &  36.3/19.6 \\
\bottomrule
%
\end{tabular}
}
\end{minipage}
\end{center}
\label{t:time}
 \end{table*}

The distribution  in the $\LX$--$z$ plane
of the 134 MCXC-II clusters with $0.4<z<0.6$   is
shown in Fig.~\ref{f:lxz}. Different flux levels are indicated by dotted lines. These trace the [0.1-2.4]~keV band flux, taking into account the $K_z$ correction for a typical gas temperature of 5~keV. The five levels are separated by 2~dex, from $1.25\times 10^{-13}$ up to  $4\times 10^{-12}$~ergs~s$^{-1}$~cm$^2$. Surveys are distinguished by different symbols and colours.
For  clusters  appearing in several catalogues (i.e. detected in different surveys), we only indicate the input catalogue used to compute the luminosity, following the MCXC-II hierarchy described in \citet{sad24}. 

The sub-samples of clusters from the ROSAT All-Sky survey (RASS) and the Serendipitous surveys appear  quasi disjoined in the redshift range under consideration. 
There are 43 RASS clusters above a luminosity of $4\times10^{44}$ ergs~s$^{-1}$ (flux above $\simeq 10^{-12}$ ergs~s$^{-1}$~cm$^2$), from the  NORAS \citep{boh00}, REFLEX \citep{boh04} and/or  MACS catalogues. Following the MCXC-II nomenclature, the latter include  the MACS-DR1 sample of the highest-z ($z>0.5$) clusters \citep{ebe07}, the DR2 flux-limited sample of the brightest $0.3<z<0.5$ MACS clusters \citep{ebe10} and  the deeper MACS-DR3 catalogue \citep[][excluding overlap with DR1 and DR2]{man12}\footnote{MCXC-II also includes additional MACS clusters recovered from non-catalogue studies (the {\tt MACS\_MISC} sub-catalogue). We  discarded them in this study as their X-ray selection function is not defined. In the redshift range under consideration, this concerns four clusters from the on-going extension of the MACS survey in the South  \citep{rep18}.}.
Only six RASS clusters lie below $4\times10^{44}$ ergs~s$^{-1}$, which are clusters from the North Ecliptic Pole (NEP) deeper part of the All-sky survey \citep{hen06}.
In contrast, all 80 clusters  from ROSAT Serendipitous  surveys  lie at   $\LX <\,10^{44}$ ergs~s$^{-1}$, with the exception of  RXJ~1120.1+4318 at z=0.6  \citep{rom00}. These come from the 160SD \citep{vik98,mul03}, 400SD \citep{bur07}, B-SHARC  \citep{rom00}, S-SHARC \citep{bur03},  WARPS-I \citep{per02} and WARPS-II \citep{hor08} catalogues. 

Finally there are five EMSS  \citep{gio90,hen04} clusters that were not rediscovered in the subsequent ROSAT surveys, three at luminosity below  $4\times10^{44}$ ergs~s$^{-1}$ and two above. The two other EMSS clusters in the redshift range are the luminous clusters  MCXC J0018.5+1626 (CL0016+16 at $z=0.55$) and  MCXC J0454.1-0300 ($z=0.54$), rediscovered in the MACS survey.
Hereafter, the  high-L$_{X}$ and low-L$_{X}$ subsamples include clusters above and below a luminosity of $4\times10^{44}$~ergs~s$^{-1}$, respectively.

\subsection{\excpres\ sample}

We searched for  \xmm\ observations, with pointing position within 5\arcmin\ of the cluster centre,  available by  September 2021 in the ESA  archive\footnote{\url{https://www.cosmos.esa.int/web/xmm-newton/xsa}}.  The observed clusters  are marked with red circles in Fig.~\ref{f:lxz} and those selected to form the \excpres\ sample are marked with red boxes. 

The core of the \excpres\ sample is the \xmm\ Large-Programme follow-up of EMSS, NORAS, REFLEX, B-SHARC, S-SHARC, 160SD and WARPS-I samples (programme ID $\#030258$ with re-observation of flared observations in  ID $\#040275$ and $\#050243$, combined with previous archival data). It was designed to study evolution from comparison with \rexcess\ data, similarly constructing  a representative sample of clusters centered at a median redshift of $z=0.5$  with homogenous coverage of the luminosity space \citep{arn08}.  The redshift excursion below $\Lv \sim 4\times10^{44}$\,ergs~s$^{-1}$ was set to $0.45<z<0.55$ and increased to $0.4<z<0.6$ to include all the rare high $z$ massive clusters known at that time.  Additional observations essentially extend the high luminosity coverage, mostly thanks to the publication of MACS clusters and corresponding follow-up.    

\begin{table*}[!t]
\caption{ \footnotesize Physical parameters of the sample. Columns 1-3: Cluster index, MCXC name and redshift. 4: Hydrogen column density from the LAB survey 21~cm measurements \citep{lab}. Columns 5-6: $\MV$ is the mass within  $\RV$, the radius  enclosing  500 times the critical density at the cluster redshift. $\TX$ is the temperature measured in the $[0.15$--$0.75]\,\RV$ aperture.  $\MV$ and $\TX$ are  measured  iteratively using the $\MV$--$\YX$  relation \citep{arn07} assuming self-similar evolution, where  $\YX$ is  the product of $\TX$ and the gas mass within $\RV$. 7-8: Count rate in the $[0.3-2]$~keV energy band within an aperture of $\RV$ in radius, and corresponding signal-to-noise (S/N) ratio.
}
\label{tab:global}
\begin{center}
\begin{tabular}{llcccrrrr}
\toprule
Obs. & Name &  z  & $N_{\rm H}$          & $T_{\rm X}$ & $\MV$       & $CR_{500}$  & $ {\rm S/N}_{500} $\\
     &      &     & $10^{20}$\,cm$^{-2}$ & keV         &  $10^{14}$\msol &  ct/s  &   \\
\midrule
  1  &   MCXC J0018.5+1626 & $0.541$ &  $ 3.99$ & $  9.66\pm0.33$ & $   9.69\pm0.23$ & $  1.17$ & $149$ \\
  2  &   MCXC J0030.5+2618 & $0.500$ &  $ 3.71$ & $  5.40\pm0.42$ & $   3.12\pm0.18$ & $  0.20$ & $ 47$ \\
  3  &   MCXC J0221.1+1958 & $0.450$ &  $ 9.12$ & $  5.44\pm0.41$ & $   2.94\pm0.16$ & $  0.24$ & $ 50$ \\
  4  &   MCXC J0257.1-2326 & $0.505$ &  $ 2.08$ & $  8.32\pm0.40$ & $   7.30\pm0.24$ & $  1.10$ & $102$ \\
  5  &   MCXC J0417.5-1154 & $0.443$ &  $ 3.32$ & $  9.01\pm0.36$ & $  11.75\pm0.45$ & $  3.29$ & $117$ \\
  6  &   MCXC J0454.1-0300 & $0.539$ &  $ 3.92$ & $  9.06\pm0.32$ & $   8.18\pm0.20$ & $  1.13$ & $122$ \\
  7  &   MCXC J0522.2-3624 & $0.472$ &  $ 3.63$ & $  4.25\pm0.42$ & $   2.21\pm0.17$ & $  0.16$ & $ 34$ \\
  8  &   MCXC J0647.8+7014 & $0.591$ &  $ 8.53$ & $  8.03\pm0.19$ & $   7.39\pm0.12$ & $  0.79$ & $156$ \\
  9  &   MCXC J0717.5+3745 & $0.546$ &  $ 6.63$ & $ 10.30\pm0.15$ & $  12.83\pm0.17$ & $  1.53$ & $359$ \\
 10  &   MCXC J0856.1+3756 & $0.411$ &  $ 3.21$ & $  6.20\pm0.42$ & $   4.89\pm0.24$ & $  0.59$ & $ 54$ \\
 11  &   MCXC J0911.1+1746 & $0.505$ &  $ 3.28$ & $  6.70\pm0.28$ & $   5.48\pm0.22$ & $  0.55$ & $ 99$ \\
 12  &   MCXC J0943.1+4659 & $0.407$ &  $ 1.00$ & $  5.53\pm0.17$ & $   4.93\pm0.11$ & $  0.70$ & $100$ \\
 13  &   MCXC J0957.8+6534 & $0.530$ &  $ 5.33$ & $  2.86\pm0.31$ & $   1.05\pm0.08$ & $  0.07$ & $ 31$ \\
 14  &   MCXC J1003.0+3254 & $0.416$ &  $ 1.68$ & $  3.71\pm0.25$ & $   2.21\pm0.11$ & $  0.35$ & $ 40$ \\
 15  &   MCXC J1120.1+4318 & $0.600$ &  $ 2.97$ & $  5.00\pm0.32$ & $   3.53\pm0.15$ & $  0.35$ & $ 54$ \\
 16  &   MCXC J1149.5+2224 & $0.544$ &  $ 1.92$ & $  8.79\pm0.60$ & $   9.36\pm0.63$ & $  1.14$ & $ 47$ \\
 17  &   MCXC J1206.2-0848 & $0.441$ &  $ 4.35$ & $ 10.15\pm0.32$ & $  10.83\pm0.25$ & $  2.34$ & $203$ \\
 18  &   MCXC J1244.0+1653 & $0.549$ &  $ 1.77$ & $  4.60\pm0.24$ & $   2.86\pm0.12$ & $  0.48$ & $ 85$ \\
 19  &   MCXC J1311.5-0551 & $0.461$ &  $ 2.43$ & $  5.69\pm0.99$ & $   2.38\pm0.28$ & $  0.09$ & $ 23$ \\
 20  &   MCXC J1347.5-1144 & $0.452$ &  $ 4.60$ & $ 11.74\pm0.25$ & $  12.01\pm0.17$ & $  4.80$ & $327$ \\
 21  &   MCXC J1411.4+5212 & $0.462$ &  $ 1.32$ & $  4.88\pm0.33$ & $   3.26\pm0.23$ & $  0.81$ & $ 86$ \\
 22  &   MCXC J1423.7+2404 & $0.543$ &  $ 2.20$ & $  5.90\pm0.25$ & $   4.64\pm0.18$ & $  1.01$ & $128$ \\
 23  &   MCXC J1623.5+2634 & $0.427$ &  $ 3.12$ & $  5.73\pm0.96$ & $   4.63\pm0.59$ & $  0.21$ & $ 24$ \\
 24  &   MCXC J1701.3+6414 & $0.453$ &  $ 2.28$ & $  4.00\pm0.19$ & $   2.91\pm0.13$ & $  0.37$ & $ 66$ \\
 25  &   MCXC J2129.4-0741 & $0.589$ &  $ 4.32$ & $  8.28\pm0.79$ & $   7.39\pm0.66$ & $  0.77$ & $ 47$ \\
 26  &   MCXC J2146.0+0423 & $0.531$ &  $ 4.82$ & $  5.16\pm0.59$ & $   2.49\pm0.20$ & $  0.13$ & $ 31$ \\
 27  &   MCXC J2214.9-1400 & $0.503$ &  $ 2.88$ & $  8.19\pm0.48$ & $   7.95\pm0.44$ & $  1.17$ & $ 83$ \\
 28  &   MCXC J2228.6+2036 & $0.412$ &  $ 4.26$ & $  8.16\pm0.30$ & $   8.73\pm0.23$ & $  1.66$ & $133$ \\
 29  &   MCXC J2243.3-0935 & $0.444$ &  $ 4.02$ & $  7.44\pm0.09$ & $   8.67\pm0.10$ & $  1.69$ & $302$ \\
 30  &   MCXC J2328.8+1453 & $0.497$ &  $ 3.88$ & $  3.17\pm0.25$ & $   1.31\pm0.08$ & $  0.05$ & $ 38$ \\
 31  &   MCXC J2359.5-3211 & $0.478$ &  $ 1.18$ & $  3.57\pm0.33$ & $   1.79\pm0.12$ & $  0.13$ & $ 37$ \\
\bottomrule 
\end{tabular}
\end{center}
 \end{table*}

The high-L$_{X}$ subsample has good \xmm\ follow-up coverage.
The EMSS and REFLEX/NORAS clusters have all  been observed with deep \xmm\ pointings (nine clusters), as well as the only cluster from a ROSAT serendipitous survey (a B-SHARC object). The \xmm\ follow-up of the MACS clusters depends on the sub-catalogues. The \xmm\ follow-up of the MACS-DR1 and MACS-DR2 clusters is nearly complete, with twelve clusters observed. Only one MACS-DR1 cluster (MCXC J0025.4-1222 at $z=0.584$)  and three DR2 clusters (MACS J0152.5-2852, MACS J0159.8-0849, MACS J0358.8-2955) are lacking observations. In contrast, only three of the seventeen DR3 clusters have been observed. 
We decided to  discard these three observations and to include the remaining  22 clusters observed by \xmm. In this way, above a luminosity of $4\times10^{44}$ ergs~s$^{-1}$, our sample constitutes a nearly complete follow-up, with a completion factor of 22/26 or 85\%,  of cluster catalogues (EMSS, NORAS/REFLEX,  MACS-DR1/DR2 and B-SHARC), with a well defined selection function. 

The  \xmm\ follow-up at lower luminosity is  more sparse. Beyond the objects of the LP observations mentioned  above, no archival observations are deep enough to allow anything but a global temperature measurement to be obtained. This is discussed in more detail in Appendix\,\ref{ap:discard}. The final selection at low luminosity includes nine clusters in the $0.44<z<0.56$ redshift range with  $\Lv > 10^{44}$\,ergs~s$^{-1}$. In this redshift range, four objects with archival \xmm\ data are not included: one is actually a point source, and three are SHARC clusters with insufficiently deep observations. 

The final \excpres\ sample comprises 31 clusters in three boxes in the $\Lv$--$z$ plane (see  Fig.~\ref{f:lxz}),  centred at a
redshift of $z=0.5$, with an approximately equal number of clusters in three equal logarithmically spaced luminosity bins. There are nine clusters in the low-luminosity bin $10^{44}<\Lv<4\times10^{44}$\,ergs~s$^{-1}$, nine clusters at intermediate luminosity, $4\times10^{44}<\Lv<12\times10^{44}$\,ergs~s$^{-1}$,   and thirteen in the high luminosity bin, $12\times10^{44}<\Lv<36\times10^{44}$\,ergs~s$^{-1}$.

\begin{figure*}%
\centerline{\includegraphics[width=0.8\textwidth]{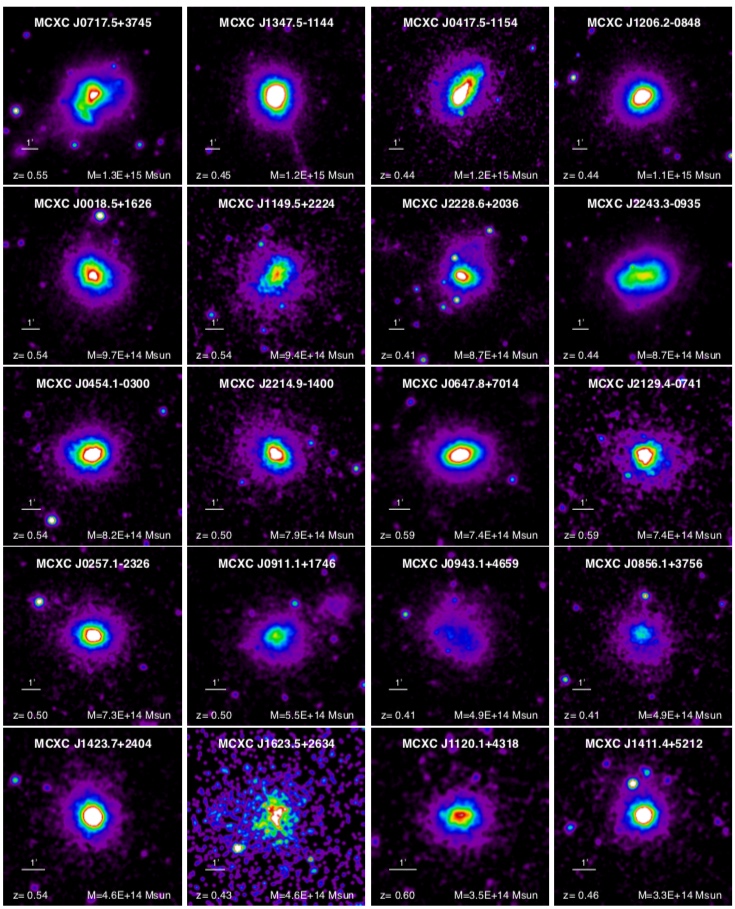}}
\caption{\footnotesize Gallery of the \xmm\ images, extracted in the $[0.3$--$2]$~keV energy band, for the 31 clusters of the sample, ordered by decreasing mass.
 Image sizes are $3 \theta_{500}$ on a side, where 
 $\theta_{500}$ is estimated from the $\MV$--$\YX$ 
 relation. Images are corrected for surface brightness dimming with $z$, divided by the emissivity in the energy 
 band, taking into account galactic absorption and instrument response, and scaled according to the self-similar model. The colour table is the same for all 
 clusters, so that the images would be identical if clusters obeyed self-similarity.}
\label{f:gal1}
\end{figure*}

\begin{figure*}%
\centerline{\includegraphics[width=0.9\textwidth]{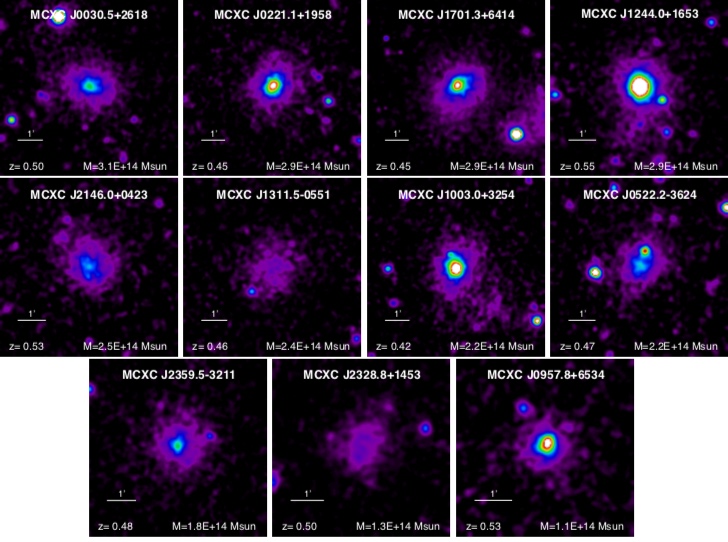}}
\caption{\footnotesize Gallery of X-ray images for the clusters of the sample, continued.}
\label{f:gal2}
\end{figure*}

\section{\xmm\ observations and data processing}
\label{s:obs} 
\subsection{Data preparation}

All data sets were retrieved and reprocessed with the \xmm\ Science Analysis Software, using the methods described extensively in for instance \citet[][]{bar17}.  In brief, the events list were (i) cleaned for solar flare contamination \citep{pra03}; (ii) filtered with PATTERN-selected (0-12 for EMOS and 0-4 for EPN); (iii) corrected for vignetting (by attributing a vignetting weight function to each event, see \citealt{arn01}); (iv) point source subtracted (after detection in the [0.3-2]~keV and [2-5]~keV  co-added EMOS+EPN image using the SAS task \verb+ewavedetect+, tuned to a detection threshold of 5$\sigma$ and double-checked visually).  All clusters have effective observation from the three \xmm\ cameras, except MS1621.5+2640 whose PN data had to be discarded due to a high rate of contamination by solar flares. Table~\ref{t:time} lists the observation details for the sample. Exposure times are the sum of EMOS (EMOS1 and EMOS2) and the EPN effective exposure time (i.e. after flare cleaning).

The (solar) particle plus instrumental background templates used in
later spatial and spectral analysis were obtained by stacking Filter
Wheel Closed (FWC) observations for each camera. The same
cleaning, PATTERN selection and vignetting correction steps were applied
to the FWC events lists. The FWC files were finally cast to match
the astrometry of each cluster observation and renormalised to the
quiescent count rate in the [10-12] and [12-14]~keV bands for
EMOS and EPN, respectively.

\subsection{Imagery and spectral analysis}

Images and surface brightness ($S_{\rm X}$) profiles  were extracted  in the $[0.3 - 2.0]$~keV band in order to maximise the S/N. The $S_{\rm X}$ data points were defined in fixed {circular} radial bins of $3\farcs3$ width, hence assuming spherical symmetry.  

\begin{figure*}[]
\includegraphics[width=0.32\textwidth]{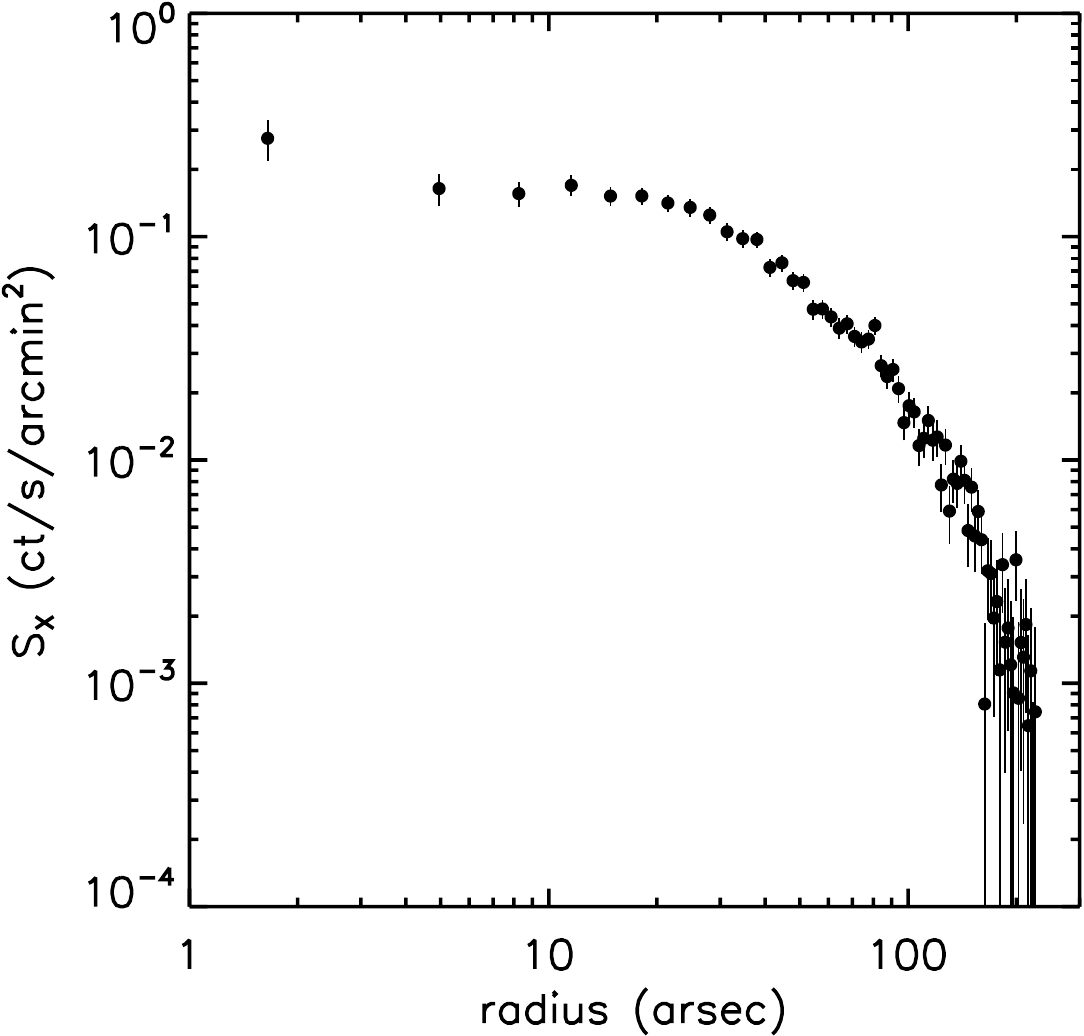}
\hfill
\includegraphics[width=0.31\textwidth]{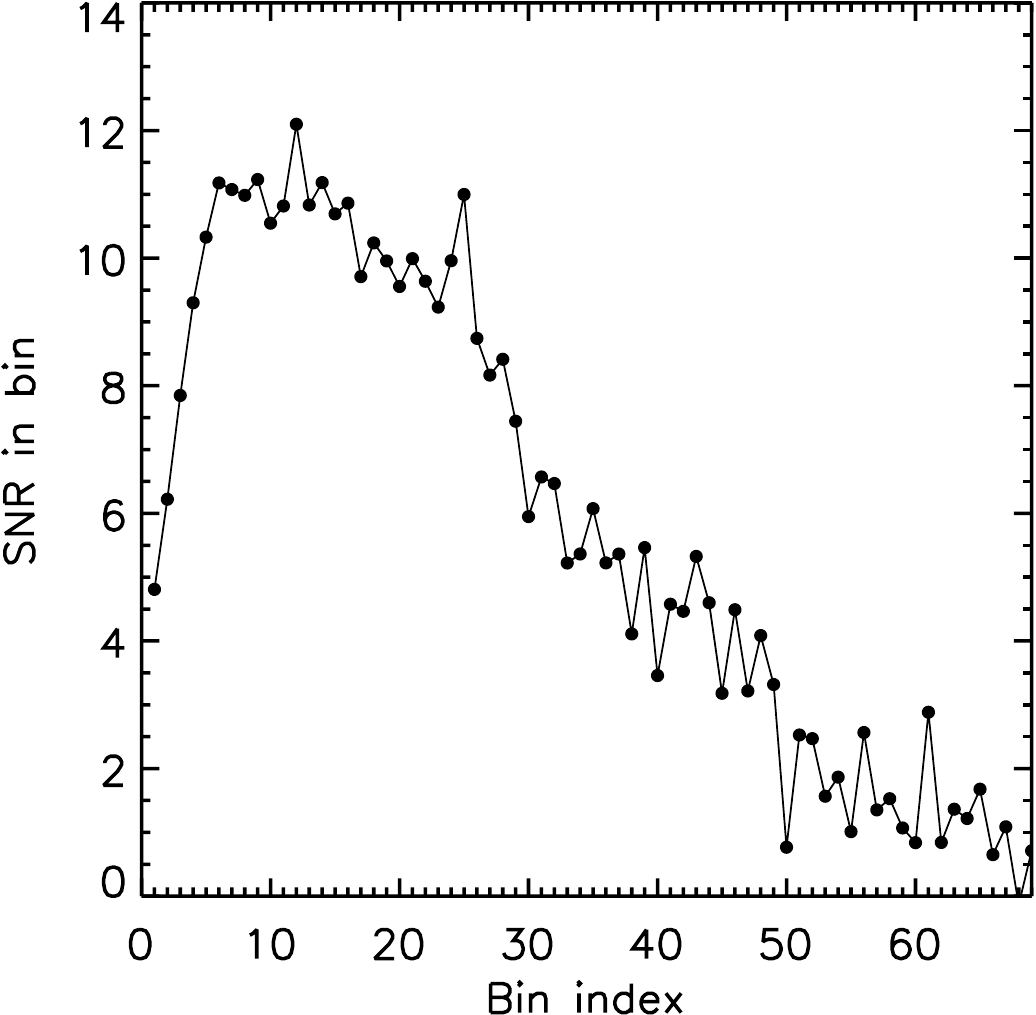}
\hfill
\includegraphics[width=0.34\textwidth]{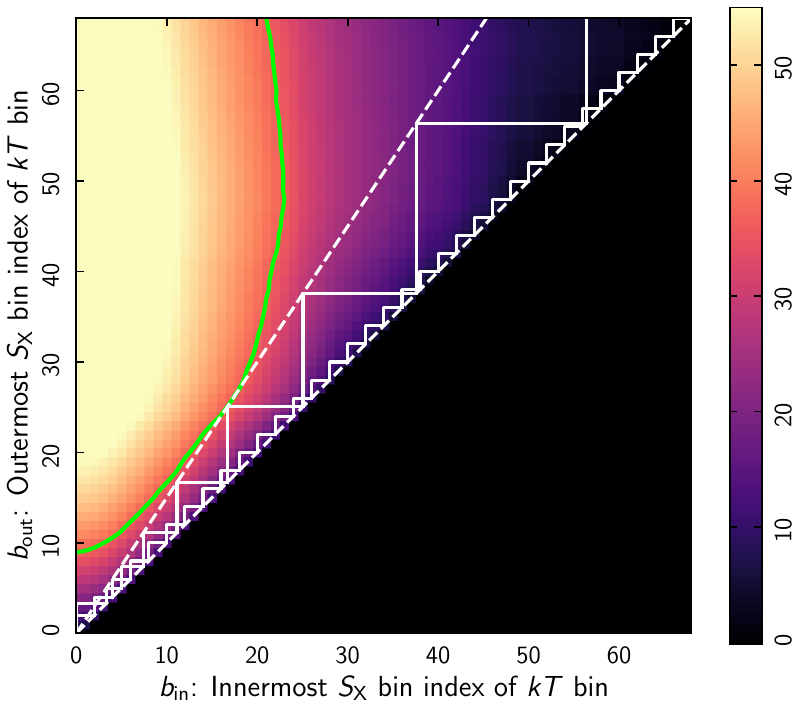}
\caption{\footnotesize 
Visual representation of the problem of binning surface-brightness  ($S_{\rm X}$) bins into temperature-profile ($kT$) bins. \emph{Left:}  $S_{\rm X}$ profile of RXJ0856.1+3756 in the [0.3-2]~keV band, extracted in 3\farcs3 {circular} radial bins. \emph{Centre:} Signal-to-noise ratio (S/N) in each bin of the $S_{\rm X}$ profile as a function of the bin index. \emph{Right:} Innermost vs outermost $S_{\rm X}$ bin indexes, overplotted on the S/N obtained from placing the range of $S_{\rm X}$ bins indicated by the axes into one single $kT$ bin. 
{The green line delineates a $30\sigma$ S/N requirement.}
Two $S_{\rm X}$ example binning schemes are also shown (solid step curves). These correspond  to a fixed binning of two input $S_{\rm X}$ bins (small fixed width steps), and a logarithmic binning of the $S_{\rm X}$ profile (continuously increasing step size). 
}
\label{f:ann1}
\end{figure*}

Spectral analysis followed that described extensively in \citet{pra10} and \citet{bar17}. Spectra were extracted in various regions of interest from the weighted events lists. The instrumental background was subtracted using the FWC spectrum from the same region in detector coordinates, normalised in the high-energy band. The cosmic X-ray background was obtained from modelling an FWC-subtracted region external to the cluster emission,  normalised in the high energy band (accounting for chip gaps, missing pixels, etc)\footnote{Typical cluster angular radii, $R_{\textrm 500}$, at these $z$ are $\lesssim 5\arcmin$.}. The model used was the sum of an unabsorbed thermal model (local bubble) added to the absorbed sum of a thermal (Galactic halo) and power law (CXB) models. It is considered constant across the cluster area and geometrically scaled in normalisation to each annulus in the modelling step. The cluster spectra were fitted with an absorbed redshifted thermal model (\verb|mekal| under XSPEC), together with the CXB model above, scaled to the area of the extraction region.
Fits were undertaken in the $[0.3-10]$~keV band. The spectrum from the $[0.15-0.75]\, R_{500}$ region was fitted with the $n_{\rm H}$ free and the result was compared to the standard Leiden/Argentine/Bonn \citep[LAB;][]{lab} 21~cm survey value. In no cases was the fitted nH significantly different from the LAB result, so this value was used.

\subsection{Global cluster properties}

The values for $\MV$ and the corresponding $\RV$ were 
computed iteratively from the $\MV$--$\YX$  relation, calibrated by \citet{arn07} 
using HE mass estimates of local relaxed clusters. We assumed that the $\MV$--$\YX$ relation obeys self-similar evolution. 
The quantity $\YX$ is defined as the product of the temperature $\TX$ measured in the $[0.15$--$0.75]\,\RV$  region 
and the gas mass within $\RV$ \citep{kra06}. The gas masses were computed from the density profiles, derived from non-parametric deprojection of surface brightness profiles and the emissivity profile computed from the 2D temperature profile, as described in \citet{cro06}. The temperature $\TX$ and mass $\MV$ for each cluster and associated errors are reported in Table~\ref{tab:global}. The table also lists the total count rate in the [0.3 -- 2.0]~keV band within an aperture of $\RV$ in radius, and the total S/N within the $\RV$ aperture. 

Figure~\ref{f:Mz} shows the distribution of the \excpres\ clusters in the $z$--$\MV$ plane. It confirms that the sample covers a similar mass range to \rexcess\, with good mass sampling. However, \excpres\ extends to slightly higher masses. Four objects have masses $\MV>10^{15}$ \msol, which is $20\%$ larger than the maximum mass of the \rexcess\ sample, which is limited by the local Universe volume.

\subsection{Images}
\label{s:images}

We computed images in the [0.3-2.0]~keV energy band in order to maximize
the S/N. The images, for each
available detector, were generated from the flare
cleaned events lists,  before point source subtraction, but  corrected
from the vignetting effect (with the SAS task \verb+evigweight+). They have been excised individually for bad
pixels and detector gaps, and then co-added. A count
image was extracted from the EPN out-of-time events lists and
subtracted from the EPN count image.

An effective exposure time image was obtained from the sum of individual detector exposure maps (outputs of SAS task \verb+eexpmap+ without vignetting correction),  weighted by the relative efficiency of each detector in the [0.3-2.0]~keV band. The total count image is then divided by this exposure map and then  corrected for surface brightness dimming with $z$, divided by the EPIC emissivity in the energy 
 band, taking into account galactic absorption and EPIC instrument response. This final image is a map  of the emission measure along the line of sight. It is then scaled according to the standard self-similar model ($EM \propto \RV h(z)^2$), so that all images would be identical for a perfect self-similar model. 
 
 The gallery of images for our 31  clusters is shown in
Figs.~\ref{f:gal1} and~\ref{f:gal2}, ordered by decreasing mass. They are displayed with a linear scale. It can be seen that the clusters cover a wide range of luminosity and morphologies.

\section{An optimal temperature profile binning method}

Temperature profile binning is commonly undertaken by simply imposing a fixed bin width, or by applying some mathematical first-principle closed-form function (e.g. logarithmic binning), or by sizing the bins to obtain a given total number of counts in a given energy band.
These solutions can lead to a suboptimal use of the data set by not accounting for the underlying signal and noise distributions. To exploit the spatial resolution of the instrument to its maximum, so that surface-brightness ($S_{\rm X}$) binning will yield sufficient counts per $kT$ bin to build and model a spectrum,  
in the following, we describe an optimal binning algorithm based on the well-known combinatorial-optimisation algorithm `dynamic programming' \citep{art07,clr}.

\begin{figure*}[!t]
\includegraphics[width=0.475\textwidth,interpolate=false]{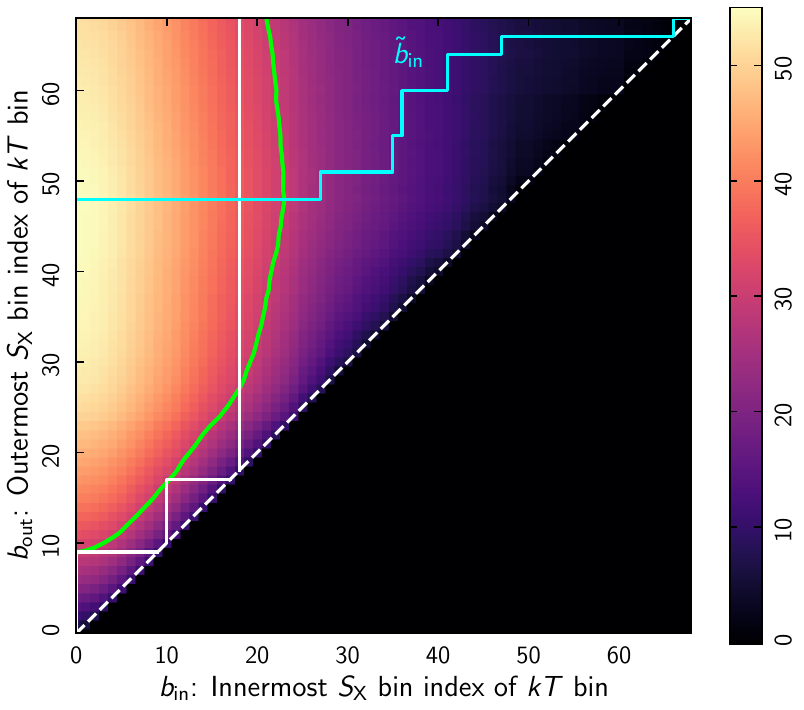}\hspace{0.5cm}
\includegraphics[width=0.475\textwidth,interpolate=false]{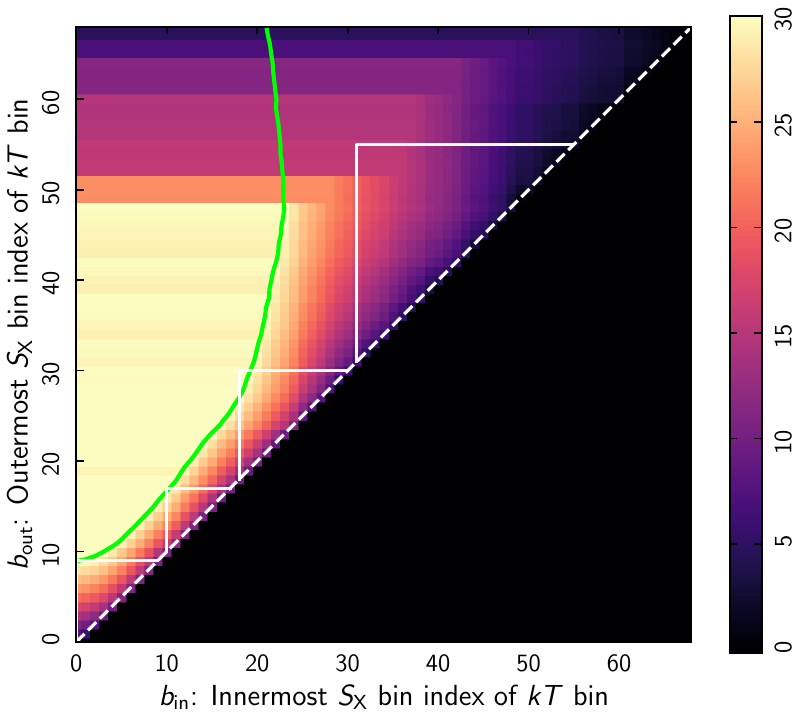}

\caption{\footnotesize Example binning solutions. \emph{Left:} Initial binning solution for the galaxy cluster RXJ0856.1+3756 (solid white steps) from the algorithm specified in Sect.~\ref{s:optimal-binning-algorithm}. 
The solid green curve delineates a $30\sigma$ S/N requirement. 
It shows that indefinitely adding more input $S_{\rm X}$ bins to an output $kT$ bin is suboptimal. The cyan steps represent  $\Tilde{b}_{\rm in}$, corresponding to the optimal innermost $S_X$ bin to include per $kT$ bin, described in the galaxy-cluster specific amendment in  Sect.~\ref{s:binning-amendment}. 
\emph{Right:} Same but with the galaxy-cluster specific amendment applied. 
Here, the S/N landscape is capped (1) globally to the S/N value at the rightmost $b_{\rm in}$ above $30\sigma$ for each $b_{\rm out}$, and (2) at $b_{\rm in} < \Tilde{b}_{\rm in}$ to the S/N value at $\Tilde{b}_{\rm in}$ for each $b_{\rm out}$.
The colour scheme has been adjusted so that the effect of the S/N landscape capping procedure is visible. }

\label{f:ann-contour}
\end{figure*}

\subsection{Problem description}

The left-hand panel of Fig.~\ref{f:ann1} shows the [0.3-2] keV surface brightness profile of RX\,J0856.1+3756 extracted in circular $3\farcs3$ radial bins. This bin width was chosen as it is an integer multiple (3) of the \xmm\ EPIC pn pixel size, and therefore maximises the angular resolution of the resulting  $S_{\rm X}$ profile in view of the mirror point spread function. The middle panel of Fig.~\ref{f:ann1} shows the resulting S/N of each $S_{\rm X}$ bin as a function of bin index. The S/N is simply defined as the ratio of the $S_{\rm X}$ to its statistical error, corresponding to $S/N=\sqrt{N}$ in a pure Poisson regime, where $N$ is the number of counts in a given bin. In our case the statistical uncertainties include the errors added in quadrature from subtraction of the instrumental (CLOSED) and astrophysical local backgrounds  \citep[see e.g.][]{bar17}. The S/N profile exhibits a typical behaviour where the S/N rises sharply to a peak in the inner regions, before tapering off quasi-linearly with increasing bin index.  
Each individual S/N is clearly insufficient to build a spectrum, hence the need for a specific rebinning of the $S_{\rm X}$ profile in order to extract the  temperature profile.

The right-hand panel of Fig.~\ref{f:ann1} illustrates the binning problem visually. 
With the input $S_{\rm X}$ bins labelled with increasing numbers outwards, the x-axis indicates the innermost $S_{\rm X}$ bin of an output $kT$ bin, while the y-axis indicates the outermost $S_{\rm X}$ bin. The underlying colour image indicates the S/N obtained from placing the range of $S_{\rm X}$ bins indicated by the axes into one single $kT$ bin\footnote{The bottom-right triangular half of the figure where $\mathit{innermost\ S_{\rm X}\ bin} > \mathit{outermost\ S_{\rm X}\ bin}$ is, obviously, to be disregarded.}.

A binning solution (a ‘partition’ hereafter) is then a set of adjacent $kT$ bins that cover the entire range of $S_{\rm X}$ bins. In Fig.~\ref{f:ann1}, a partition is represented as a contiguous set of steps starting at (0, 0) and ending at the top-right corner. A step running from innermost $S_{\rm X}$ bin $a$ to outermost $S_{\rm X}$ bin $b$ represents an output $kT$ bin covering input $S_{\rm X}$ bins from $a$ to $b$. 

Figure \ref{f:ann1} illustrates two possible partitions. The first is a  partition defined by mapping two  $S_{\rm X}$ bin into one $kT$ bin, which is represented as a series of two $S_{\rm X}$-bin tall steps running along the diagonal. The second is a logarithmic binning solution, which results in steps whose vertices lie along some straight line starting from the origin and having slope $> 1$. From such a visualisation, we would like to solve the binning problem in a data-driven way by defining a partition scheme accounting for the expected S/N level. That is, we would like the bin vertices to lie on coloured pixels having as high S/N as possible.

\subsection{Definition of optimal binning}

What does it mean to bin optimally?
In our case, optimal binning (for temperature profile measurement) is to find a way to distribute the data signal and noise in such a way that the resulting binning scheme enables the best temperature estimation, given the characteristics of the data set in question. There are, in fact, only a finite (albeit large) number of ways of binning the input $S_{\rm X}$ data into output $kT$ bins. The exact number of solutions is $2^n$, where $n+1$ is the number of input $S_{\rm X}$ bins.\footnote{We can think of the problem of partitioning $(n + 1)$ ${S_{\rm X}}$ input bins into $(r + 1)$ $kT$ output bins as choosing $r$ amongst the $n$ $S_{\rm X}$-bin boundaries as $kT$-bin boundaries.  The number of ways to do so is then $\Sigma_{r=0}^{n}\; {_nC_r}$, where ${_nC_r} = n! / r! / (n - r)!$ are the binomial coefficients appearing on the $n$-th row of Pascal's Triangle, whose sum is $2^n$.}
However, there are close to 300 $S_{\rm X}$ bins in a typical data set of ours, so the number of possible partitions can be close to $2^{300}$ or $10^{30}$, which is intractable to compute one by one. To solve this problem we have developed an algorithm based on `dynamic programming'~\citep[see e.g.][chap.~15]{clr}, which effectively considers all possible solutions by recursively building up the optimal solution, given some algorithmic criteria.

\subsection{Optimal binning algorithm}
\label{s:optimal-binning-algorithm}

\subsubsection{Fundamentals}
We want our temperature profile measurements to have the best-possible distribution of S/N across the full radial range, given the characteristics of the data set in question. Generally, increasing the S/N of one $kT$ bin implies a decrease in the S/N of an adjacent $kT$ bin. Therefore, we will want to maximise the lowest S/N of the temperature profile annuli. Given the steep drop of cluster X-ray emission with radius, the lowest S/N temperature bin is almost always the one farthest from the centre. Conversely, in the centre, where the signal is strong, we prefer to split high-S/N $kT$ bins into more data points having a S/N above some given requirement, rather than accumulating more signal into a single bin. 
We can formulate the two preferences above algorithmically as follows:
\begin{quote}
Given two partitions, A and B, of the same $S_{\rm X}$-bin range into disjoint $kT$ bins (a.k.a. subsets):
\begin{enumerate}
\item The partition whose lowest subset-S/N is higher is preferable.  When making this comparison, we cap subset S/Ns at the nominal requirement, thus treating all subsets with S/N above the requirement as equal. 
\item If the lowest subset S/N of partitions A and B are identical, the partition whose second-lowest subset-S/N is higher is preferable.
\item This process is continued.
\item If all $kT$ bins of partition A have (capped) S/N  matching $kT$ bins in partition B, but B has more $kT$ bins, then B is desirable.
\end{enumerate}
\end{quote}

\begin{figure*}[!t]
\includegraphics[width=0.475\textwidth,interpolate=false]{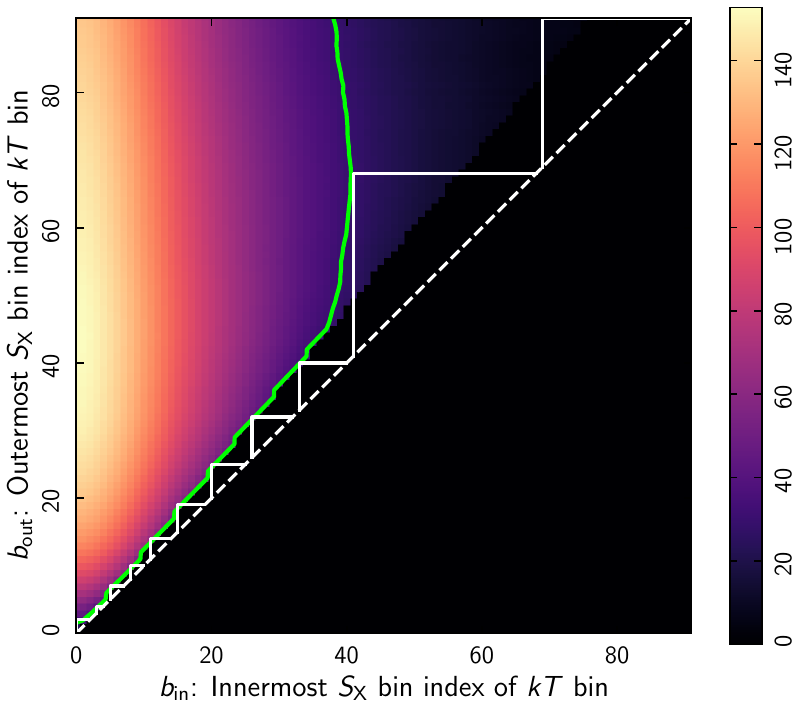}\hspace{0.5cm}
\includegraphics[width=0.475\textwidth,interpolate=false]{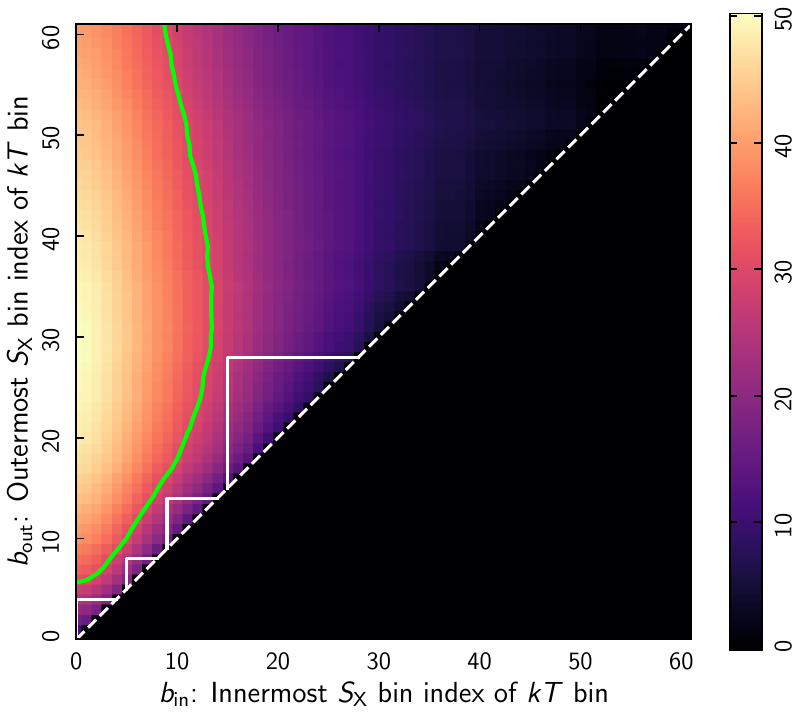}

\caption{\footnotesize Two algorithm variations. The S/N landscape is shown here uncapped for illustration, but it is capped in the binning algorithm as described in Sect.~\ref{s:binning-amendment}. In both panels, the solid curve represents a S/N of $30\sigma$. \emph{Left:} Modified S/N landscape 
and the resulting binning solution for the galaxy cluster CL0016+16. Here, the output $kT$ bins are required to fulfil four simultaneous criteria: (1) to have a target S/N of $30\sigma$ (denoted by the solid curve); (2) to have a minimum S/N $\ge 3\sigma$; (3) to include at least two $S_{\rm X}$ bins; and (4) to increase in width with a logarithmic factor of $1.2$. 
 \emph{Right:} Optimal binning for the low S/N cluster RXJ0030.5+2618, obtained from fixing the number of output $kT$ bins to four and maximising their resulting S/N (see Sect.~\ref{sec:var}). }
\label{f:ann-constrained}
\end{figure*}

This selection algorithm allows us to systematically decide between any two ways of partitioning a given $S_{\rm X}$ range of $n$ bins. With the technique of dynamic programming, the optimal binning solution for the full $S_{\rm X}$-bin range (denoted OP below) is built up recursively from the union of optimal binning solutions for shorter $S_{\rm X}$-bin ranges, applying the selection algorithm at every step. If the kT-bin boundaries defined by the optimal binning algorithm are denoted \textit{a, b, c\ldots} = \textit{0, 1, 2,... i}, the algorithm is described as follows:
\begin{enumerate}
\item We start by considering the trivial one-$S_{\rm X}$-bin subranges $[0, 0]$, $[1, 1]$, $[2, 2]$, \ldots.  The only possible (and therefore optimal), binning solution is to have one $kT$ bin containing one $S_{\rm X}$ bin: $\mathrm{OP}[a, a] = [a, a]$.
\item Then, we solve for two-$S_{\rm X}$-bin subranges $[0, 1]$, $[1, 2]$, $[2, 3]$, \ldots.  The two possible solutions are one $kT$ bin containing two $S_{\rm X}$ bins, $[a, a+1]$, or  two $kT$ bins of one $S_{\rm X}$ bin each, $[a, a] \cup [a+1, a+1]$. We use the selection algorithm described above to decide.

\item We then solve for three-$S_{\rm X}$-bin subranges $[a, a + 2]$ as the union of 'optimal solutions' for one-$S_{\rm X}$-bin and two-$S_{\rm X}$-bin subranges that we obtained above: $[a, a] \cup \mathrm{OP}[a+1, a+2]$ versus $\mathrm{OP}[a, a+1] \cup [a+2, a+2]$ versus $[a, a+2]$. Again, we use the selection algorithm to decide. We note that it suffices to consider only unions of two optimal partitions, $\mathrm{OP}[a, i] \cup \mathrm{OP}[i+1, a+2]$ for all possible $i$, and the single partition $[a, a+2]$ spanning the subrange in consideration. That is to say, we do not need to consider the solution $[a, a] \cup [a+1, a+1] \cup [a+2, a+2]$ because it is either equivalent to $[a, a] \cup \mathrm{OP}[a+1, a+2]$ and to $\mathrm{OP}[a, a+1] \cup [a+2, a+2]$, or it is not an optimal solution.

\item We then solve for four-$S_{\rm X}$-bin subranges $[a, a+3]$ similarly, by comparing unions of two optimal partitions, $\mathrm{OP}[a, i] \cup \mathrm{OP}[i+1, a+3]$ for all three possible $i$, and the single partition $[a, a+3]$.  Again, unions of more than two optimal partitions need not be considered explicitly, as they are either equivalent to some union of two optimal partitions, or are suboptimal.

\item We repeat the previous step to solve for all five-$S_{\rm X}$-bin subranges, then six-$S_{\rm X}$-bin, ..., until the n-$S_{\rm X}$-bin subrange, at which point we have the optimal binning of the full $S_{\rm X}$-bin range.
\end{enumerate}

A key characteristic of our selection algorithm is that for ${\rm Partition}\,[a, b] \cup {\rm Partition}\,[b+1, c]$ to be $\mathrm{OP}[a, c]$, the two child partitions themselves must be optimal as well.
\footnote{Suppose that $\mathrm{OP}[a, c]$ is ${\rm Partition}\,[a, b] \cup {\rm Partition}\,[b+1, c]$, but $\mathrm{OP}[a, b]$ is not ${\rm Partition}\,[a, b]$ but rather ${\rm Partition*}\,[a, b]$.  Then ${\rm Partition*}\,[a, b] \cup {\rm Partition}\,[b+1, c]$ would give higher S/N than ${\rm Partition}\,[a, b] \cup {\rm Partition}\,[b+1, c]$, contradicting the initial assertion that the latter is $\mathrm{OP}[a, c]$.}
This `optimal substructure' is what makes dynamic programming applicable to our problem.
As seen in the steps above, the recursive build-up of the final solution with dynamic programming involves computing the optimal solution for each subrange only once. These subranges are each present in an exponential number of partitions.  Thus, in effect, we are breaking apart the $2^{(n-1)}$ possible partitions of an input of length $n$ and grouping them into $n(n+1)/2$ subranges, transforming the problem from enumerating and comparing an exponential number of partitions to solving for a polynomial number of subranges.  It is therefore possible to cover all $2^{(n-1)}$ possible solutions in polynomial time.

The left-hand panel of Fig.~\ref{f:ann-contour} shows this algorithm in action, solving the binning problem in a data-driven manner for the $S_{\rm X}$ profile of RXJ0856.1+3756 from Fig.~\ref{f:ann1}. In this particular case, the optimal solution requires that the innermost $kT$ bin be larger than some outer ones; a traditional approach such as logarithmic binning is not able to produce such a solution.

It is informative to consider the roles that the partition-selection criteria play:
\begin{itemize}
    \item At the strong-signal end, the algorithm produces multiple data points by splitting $kT$ bins having S/N above a given threshold. In the left hand panel of Fig.~\ref{f:ann-contour}, it keeps these data points above, but close to, the line marking the threshold requirement.
    \item At the weak-signal end, if our sole criterion were high S/N without splitting $kT$ bins at the strong-signal end, we would always end up with one single $kT$ bin containing all data.

    This is obviously not helpful in constructing a temperature profile.
\item Finally, if we split $kT$ bins having S/N above the threshold at the strong-signal end without simultaneously imposing a S/N maximisation criterion at the weak-signal end, we would end up with the largest-radius $kT$ bin having very low S/N, typically unusable as a temperature-profile data point.
This is undesirable because 
this farthest bin is usually also the widest in a galaxy-cluster X-ray data set; discarding it would mean throwing away a large amount of expensive observational data.
\end{itemize}

\subsubsection{A galaxy-cluster specific amendment}
\label{s:binning-amendment}

While our algorithm is as generic as possible, it is not specific to typical galaxy cluster X-ray data sets. This means that it still produces suboptimal output, typically yielding outermost $kT$ bins that are too wide. It therefore requires some fine tuning to account for the characteristics of galaxy-cluster X-ray data.

The S/N pattern shown in the centre panel of  Fig.~\ref{f:ann1} is very typical of X-ray observations of galaxy clusters.
The S/N decrease with increasing radius is such that when starting from a particular innermost bin $b_{\rm in}$, it becomes counterproductive to include more $S_{\rm X}$ bins beyond some radius $b_{\rm out}$. In the example shown in the left hand panel of Fig.~\ref{f:ann-contour}, the $30\sigma$ contour line reaches a maximum at $b_{\rm in} =23$, along with a corresponding  $b_{\rm out}\approx 48$. In other words, the S/N of a $kT$ bin does not increase monotonically with the bin width. Yet the algorithm deciding between two partitions of the same $S_{\rm X}$ bin range is agnostic to this peculiar property of galaxy-cluster X-ray data, resulting in the very wide outermost $kT$ bin in the left hand panel of Fig.~\ref{f:ann-contour}, whose S/N is not as high as it might be.

We thus need to augment the algorithm to prevent it from choosing partitions containing such suboptimal $kT$ bins. In practice, this is achieved by modifying the S/N landscape shown in colour in the left hand panel of Fig.~\ref{f:ann-contour} to discourage the algorithm from choosing $kT$ bins that are too wide, that is, those containing extraneous $S_X$ bin(s), the inclusion of which \textit{lowers} the $kT$-bin S/N, as described in the last paragraph.
Because the lowest S/N $S_{\rm X}$ bins are always at large radii, 
the algorithm will expand the outermost $kT$ bin inwards as much as possible -- sometimes too much -- to maximize its S/N in the absence of such S/N landscape modification.  Instead, the optimal point to stop expanding inwards can be found by observing that if `for a given $b_{\rm in}$, there is a $b_{\rm out}$ beyond which it is counterproductive to include more $S_{\rm X}$ bins', then the corollary is `for a given $b_{\rm out}$, there is an optimal $b_{\rm in}$ beyond which it is not a good use of the overall data to include more $S_{\rm X}$ bins, despite yielding higher S/N for the $kT$ bin in consideration'.

In the following, we will call this optimal innermost $S_{\rm X}$ bin to include $\Tilde{b}_{\rm in}$. In practice, we first obtain the optimal outermost $S_{\rm X}$ bin to include by finding where the S/N reaches its maximum in each $b_{\rm in}$ column, as illustrated by the cyan steps in the left hand panel of Fig.~\ref{f:ann-contour}.  This curve also gives us $\tilde{b}_{\rm in}$ as a function of $b_{\rm out}$ by the corollary above.
To prevent the algorithm from moving further inwards, we therefore feed into the algorithm a modified copy of the S/N data, in which the values at $b_{\rm in} < \Tilde{b}_{\rm in}$ are capped to the value at $\Tilde{b}_{\rm in}$. This modified S/N landscape is shown in colour in the right hand panel of Fig.~\ref{f:ann-contour}. Now, the algorithm stops moving further inwards when it reaches $\Tilde{b}_{\rm in}$, and the resulting partitioning scheme (solid white steps) becomes optimal.

\begin{table}[!t]
\caption{\footnotesize Variations to the standard binning algorithm applied to each object in our sample. Except for the last four clusters, all binning runs have $30\sigma$ scientific S/N goal, minimum $3\sigma$ per output $kT$ bin, and minimum two input $S_{\rm X}$ bins per output $kT$ bin. Clusters with too little data to construct at least four $30\sigma$ $kT$ bins have the output $kT$-bin count fixed to four instead. The profiles for the last four clusters listed at the end of the table are from the analysis of \citet{bar19} who used a $20\sigma$ criterion. Cols. (1) and (2) ID number and name. Col. (3): Flag whether 4+ $30\sigma$ $kT$ bins is possible. Col. (4): Logarithmic binning factor. Col. (5): Optimised $kT$-bin count.
}
\begin{center}
\begin{tabular}{rlccr}
\toprule
ID & Cluster  &  Flag  & Logfac & NT\\
\midrule
  1 &    MCXC J0018.5+1626 & Y & 1.20 &   11 \\
  2 &    MCXC J0030.5+2618 & N & -- &    4 \\
  3 &    MCXC J0221.1+1958 & Y & -- &    4 \\
  4 &    MCXC J0257.1-2326 & Y & 1.20 &    9 \\
  5 &    MCXC J0417.5-1154 & Y & 1.40 &    8 \\
  6 &    MCXC J0454.1-0300 & Y & 1.20 &   10 \\
  7 &    MCXC J0522.2-3624 & N & -- &    4 \\
  8 &    MCXC J0647.8+7014 & Y & 0.00 &   10 \\
 10 &    MCXC J0856.1+3756 & Y & -- &    4 \\
 12 &    MCXC J0943.1+4659 & Y & 1.20 &    9 \\
 13 &    MCXC J0957.8+6534 & N & -- &    4 \\
 14 &    MCXC J1003.0+3254 & N & -- &    4 \\
 15 &    MCXC J1120.1+4318 & Y & -- &    4 \\
 17 &    MCXC J1206.2-0848 & Y & 1.20 &   13 \\
 18 &    MCXC J1244.0+1653 & Y & 1.20 &    7 \\
 19 &    MCXC J1311.5-0551 & N & -- &    4 \\
 20 &    MCXC J1347.5-1144 & Y & 1.20 &   12 \\
 21 &    MCXC J1411.4+5212 & Y & 1.10 &    6 \\
 22 &    MCXC J1423.7+2404 & Y & 1.20 &    8 \\
 23 &    MCXC J1623.5+2634 & N & -- &    4 \\
 24 &    MCXC J1701.3+6414 & Y & 1.10 &    7 \\
 26 &    MCXC J2146.0+0423 & N & -- &    4 \\
 28 &    MCXC J2228.6+2036 & Y & 1.20 &   11 \\
 29 &    MCXC J2243.3-0935 & Y & 1.25 &   11 \\
 30 &    MCXC J2328.8+1453 & N & -- &    4 \\
 31 &    MCXC J2359.5-3211 & N & -- &    4 \\
\midrule
 9 &    MCXC J0717.5+3745 & Y & 1.30 &   10 \\
 11 &    MCXC J0911.1+1746 & Y & 1.30 &    9 \\
  16 &    MCXC J1149.5+2224 & Y & 1.30 &    6 \\
  27 &    MCXC J2214.9-1400 & Y & 1.30 &    8 \\
\bottomrule 
\end{tabular}
\end{center}
\label{t:ann}
\end{table}

\subsubsection{Algorithm variations}
\label{sec:var}
In practice, there are sometimes extra constraints to consider. We encode such constraints as additional modifications to the input S/N landscape, 
allowing us to run the optimal binning algorithm without change.
For example, downstream data processing may require wider $kT$ bins than the input single $S_{\rm X}$ bins, which translates to the requirement of a minimum input $S_{\rm X}$-bin count $N_{\rm in, min}$ per output $kT$ bin.
We can implement this by zeroing the S/N values at pixels where 
$b_{\rm out} - b_{\rm in} < N_{\rm input,min}$
(i.e. those closest to the diagonal).
Alternatively, we may want $kT$ bins at least as wide as those one would obtain from logarithmic binning. We can satisfy this constraint by zeroing S/N values where the ratio
$b_{\rm out}/b_{\rm in}$ is less than 
the desired logarithmic binning factor.
We can also similarly impose a minimum S/N per output $kT$ bin by zeroing unacceptable pixels in the input S/N landscape. The left hand panel of Figure~\ref{f:ann-constrained} demonstrates how the algorithm operates when given a number of such constraints.

Instead of starting from a fixed S/N requirement, we can also fix the number of output $kT$ bins,  and let our optimisation algorithm maximise the lowest S/N amongst them. 
To fix the number of output $kT$-bin at $N_{\rm out, fixed}$, we first compute binning solutions to sub-problems of constant numbers of output $kT$-bin, that is $1, 2, \ldots, N_{\rm out, fixed}$.
From these solutions, we build the final fixed-size  solution through dynamic programming. This approach tends to be more applicable when the overall S/N is too low to meet an imposed S/N requirement. The right hand panel of Fig.~\ref{f:ann-constrained} shows an application of this algorithm variation.

\begin{figure}[!t]
\includegraphics[width=\columnwidth]{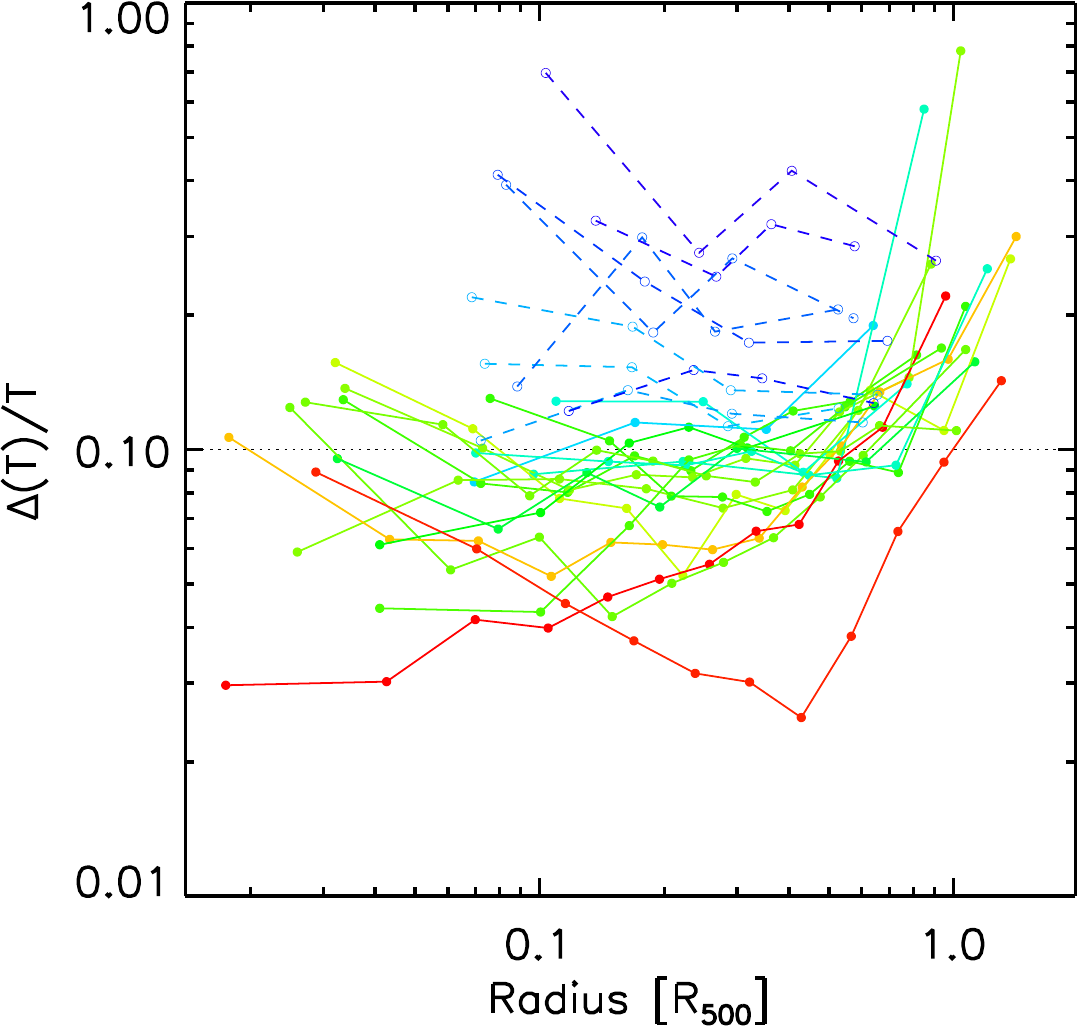}
\caption{\footnotesize Relative uncertainty on the temperature as a function of scaled radius, for the profiles defined  with $30\sigma$ scientific S/N goal (Table~\ref{t:ann}). The profiles are colour coded as a function of the total S/N in the soft-band (Table~\ref{t:time}), {from blue (low S/N) to red (high S/N).}}
\label{f:snr}
\end{figure}
\begin{figure*}[!t]
\includegraphics[width=\textwidth]{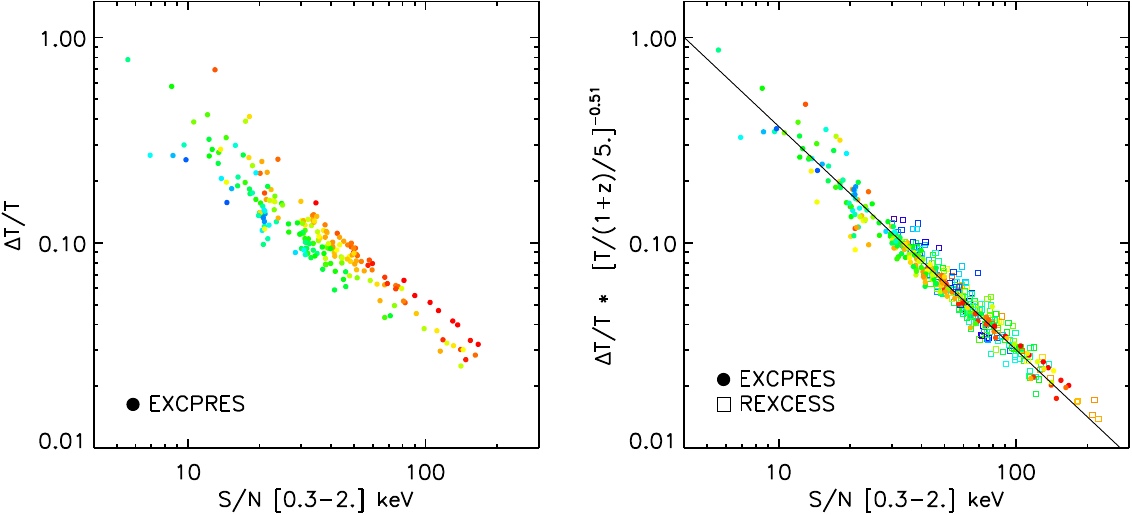}
\caption{\footnotesize Relation between S/N and fractional temperature uncertainty. {\it Left}: Relation between relative error on  the temperature and S/N. Each point corresponds to a bin of a temperature profile. {\it Right}: Rescaled relation, including data from \rexcess\ at lower redshift. Points are colour coded according to their temperature, lower and higher $kT$ being displayed in blue and red, respectively.}
\label{f:snr2}
\end{figure*}
\subsection{Implementation}
\label{sec:implementation}

For each cluster in our sample, we first produced $S_{\rm X}$ profiles in the [0.3, 2.0]~keV band with data points defined in fixed circular radial bins of $3\farcs3$ width. 
We ran our optimal binning algorithm with a S/N criterion of $30\sigma$ per $kT$ bin, which ensures a $\sim$10\% precision on the temperature measurement in each bin. Owing to their faint X-ray emission, some of the objects required a variation to the standard binning algorithm such that a minimum of four optimally binned annuli were produced, as described in Sect.~\ref{sec:var}. For high S/N objects, we imposed an additional logarithmic binning factor. We add a criterion that the $kT$ bins are always larger than one $S_{\rm X}$ bin by imposing that the  
minimum size of the $kT$ bins be at least two  $S_{\rm X}$ bins wide. Table~\ref{t:ann} indicates the variations applied to each cluster. After application of the binning criteria to the $S_{\rm X}$ profiles, spectra were extracted from each corresponding $kT$ annulus, and fitted as described above. 

The temperature profiles of four MACS clusters (ID \#9, 11, 16, and 27) listed at the end of Table ~\ref{t:ann} are from the study of {\it Planck}-selected clusters by \citet{bar19}. They used the present optimal binning procedure, but with a S/N criterion of $20\sigma$ instead. For consistency with their published work, we kept the original profiles. We also keep their analysis of a fifth MACS cluster, MCXC~J2129.4-0741 (\#25), which has a manually-defined four-bin temperature profile.

\section{Results}


\subsection{Optimally binned temperature profiles}

The individual temperature profiles are shown in Appendix~\ref{a:indt}. Figure~\ref{f:snr} shows the relative uncertainty on the temperature as a function of scaled radius for the full sample, colour-coded by the total soft-band S/N in the $\RV$ aperture (i.e. S/N$_{500}$ in Table~\ref{t:time}). One can distinguish three regimes. For very high S/N observations (red, orange), the binning out to large radius is driven essentially by the logarithmic binning factor. For observations with an intermediate S/N (yellow, green), the algorithm bins the $S_{\rm X}$ at $30\sigma$ until this is no longer feasible, followed by an additional bin out to $\RV$. The majority of such intermediate S/N clusters therefore display a flat fractional temperature uncertainty with radius, corresponding to $\Delta T/T \sim 10\%$, followed by one single bin with slightly higher fractional uncertainty at the largest scaled radius. Finally, for observations with a low S/N (blue), the algorithm maximises the common S/N for the four-bin minimum criterion; for these objects, the fractional temperature uncertainty is flat with scaled radius, but at a level greater than 10\%.

Figure~\ref{f:snr2}  shows the relative precision on each point of the temperature profile as a function of the corresponding S/N in that annular bin. There is a good correlation, justifying a posteriori the  underlying  assumption of the binning method. There is a dispersion with an apparent increased error with increasing temperature at a given S/N. As the S/N is defined in the soft band this is expected. The upper limit of the band is below the cut-off energy of the bremsstrahlung energy at $E\sim$kT, in most cases. The soft band flux (and thus the S/N in that band) is therefore  insensitive to temperature while, with increasing temperature, the energy cut-off moves to higher energy, where the instrument effective area decreases. As this sets the constraints on temperature, its precision also decreases at a given S/N. 

Adding information from \rexcess\ to increase the redshift leverage, we found an empirical power law relation between the relative error, once rescaled as a function of $T/(1+z)$, and the S/N:
\begin{equation}
    \frac{\Delta(T)}{T} \times \left[T_{5}/(1+z)\right]^{-0.51} = 0.11\,\times    \left[\frac{{\rm S/N}}{30}\right]^{-1.09}
\label{e:dtt}
\end{equation}
where $T_{5}$ is the temperature in units of 5 keV. This is illustrated on the right panel of  Fig.~\ref{f:snr2}. This relation can be used, for instance, to define the  exposure times required to reach a given temperature precision from simple information on the surface brightness profile or the overall count rate.

\subsection{Scaled temperature profiles}\label{s:model}

\subsubsection{3D temperature profiles}

The 2D temperature profiles derived using the algorithm described above were deprojected and PSF-corrected as described in \citet{bar18}.  In brief, each individual 2D profile was modelled with a 3D parametric model similar to that proposed by \citet{vik06}, convolved with a response matrix that simultaneously takes into account projection and PSF redistribution. The weighting scheme introduced by \citet{vik06b} was used to correct for the bias introduced by fitting isothermal models to multi-temperature plasma. Fitting was undertaken with Bayesian maximum likelihood estimation and Markov chain Monte Carlo (MCMC) sampling, using \verb|emcee| \citep{emcee}. The final deprojected, PSF-corrected profiles were derived from the best-fitting model temperature at the weighted radii corresponding to the 2D annular binning scheme. The left-hand panel of Fig.~\ref{f:sct} shows the resulting 3D profiles scaled in terms of $R_{500}$ and $T_{\rm X}$, the spectral temperature in the $[0.15-0.75]\,R_{500}$ region, colour-coded as a function of mass. The individual 2D and 3D temperature profiles and associated best fitting models are shown in Appendix~\ref{a:indt}.

\subsubsection{Sample average profile}

We fitted the scaled profiles $T_{\rm m}(x) = T(r)/T_{\rm X}$  with a model consisting of an analytical profile with a radially-varying intrinsic scatter term using the formalism described in \citet{pra22}: 
\begin{gather}
    T_{\rm m}(x)= f(x) 
\end{gather}
with
\begin{gather}
    x= r/R_{500},
\end{gather}

with $f(x)$ being described by the model proposed by \citet{vik06}
\begin{gather}
    f(x) = T_0 \times \frac{(y+T_{\rm  min}/T_0)}{(y+1)} \times \frac{(x/x_t)^{-a}}{\left[ 1+(x/x_t)^b\right]^{c/b}} \ \ ; \ y=\left(\frac{x}{x_{\rm cool}}\right)^{a_{\rm cool}}.
\end{gather}
 Accounting for measurement errors, the probability of measuring a given scaled gas temperature, $T$, at given scaled radius, $x$, is
\begin{eqnarray}
    p(T|x) &  = & \mathcal{N}[\log{T_{\rm m}}(x),\sigma^2(x) ]\\
    \sigma^2(x) & = & \sigma^2_{\rm int} + \sigma^2_{\rm stat}
\end{eqnarray}
where $\mathcal{N}$ is the log-normal distribution, while the variance, $\sigma^2(x)$, is the quadratic sum of the statistical error $\sigma_{\rm stat}$ on the measured $\log{T(x)}$ , and of the intrinsic scatter on the median profile $\log{T_{\rm m}(x)}$ at radius $x$, $\sigma_{\rm int}(x)$. As in \citet{pra22}, we used a non-analytical form for the intrinsic scatter, defining $\sigma_{\rm int}(x)$ at $n$ equally-spaced points in $\log(x)$ with $\sigma_{\rm int}(x)$ at other radii being computed by spline interpolation. We used $n = 7$ between $x_{\rm min} = 0.01$ and $x_{\rm max} = 1$.

\begin{figure*}[!t]
\includegraphics[width=1.09\columnwidth]{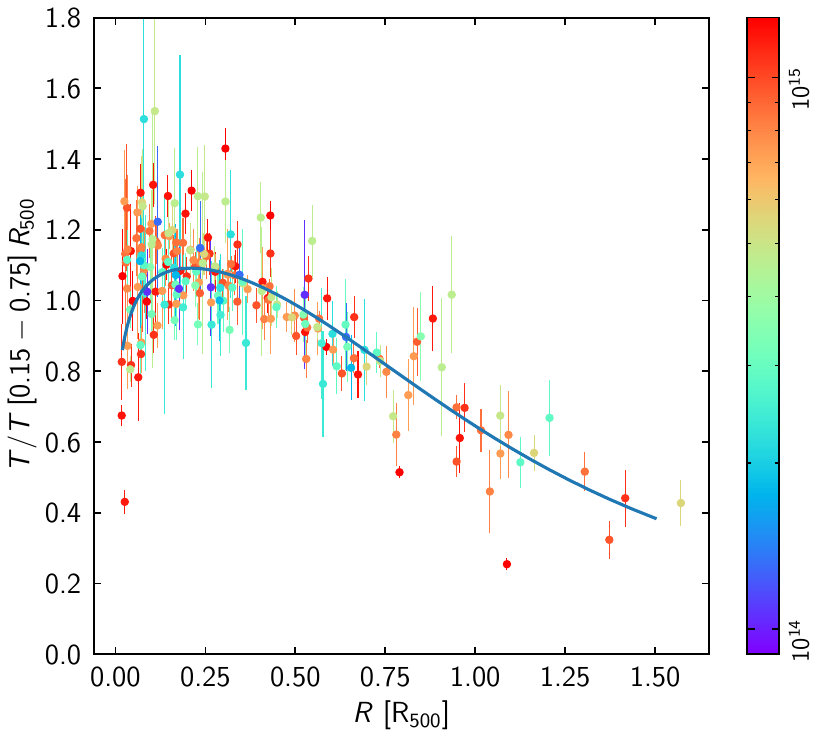}
\hfill
\includegraphics[width=0.975\columnwidth]{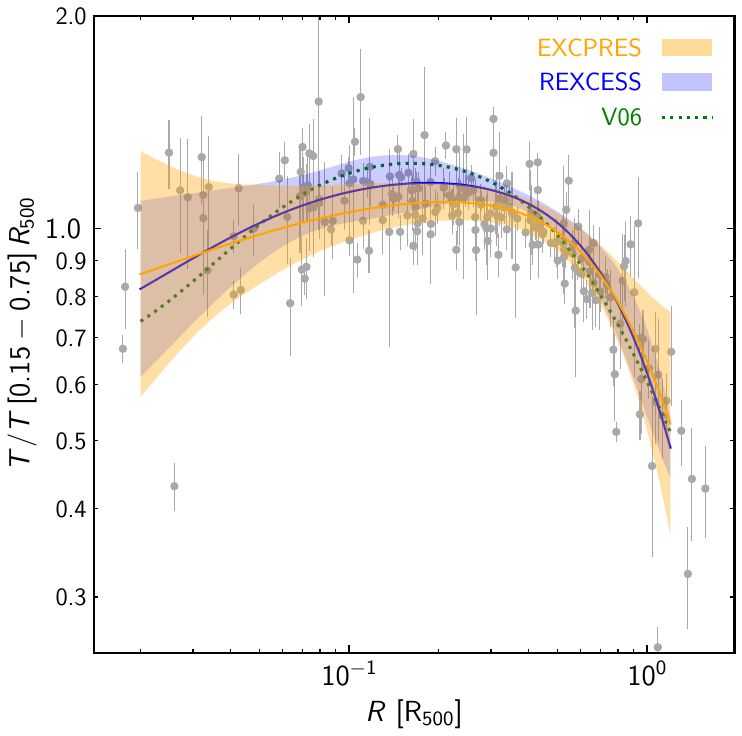}
\caption{\footnotesize {\it Left}: Deconvolved, deprojected, 3D temperature profiles of the \excpres\ sample, scaled by $R_{500}$ and the spectral temperature in the $[0.15-0.75]\,R_{500}$ region, colour-coded by total mass. The best-fitting analytical model (Eqns. 2-5) is overplotted. {\it Right}: Scaled 3D profiles of the \excpres\ sample (points with error bars)  and best-fitting model (orange), compared to the best-fitting model to the scaled temperature profile of the \rexcess\ sample (blue). Envelopes indicate the radially-varying intrinsic scatter term. The dotted line shows the best-fitting model to the local cool core sample of \citet{vik06}.}
\label{f:sct}
\end{figure*}

The likelihood of  a  set of scaled temperature  profiles measured for  a sample of  $i=1, N_{\rm c}$ clusters is:
\begin{gather}
\mathcal{L}  = \prod_{i=0}^{N_{\rm c}} \prod_{j=0}^{ N_{\rm R}[i]} p\,(T_{i, j} \vert x_{i,j}), %
\end{gather}
where   $N_{\rm R}[i]$ is the number of  points  of the  profile of cluster $i$, and the quantity $T_{ i,j}= T[i,j]/T_{\rm X_{i}}$ is the scaled temperature   measured at each scaled radius $x_{i,j}=r[i,j]/R_{500,i}$, with $T[i,j]$ and $r[i,j]$ being  the physical gas temperature and radius.  The statistical error on $\log{T_{i,j}}$ is $\sigma_{\rm stat,i,j}$. 
We  fitted the data (i.e. the set of observed $T[i,j]$ ) using a Bayesian maximum likelihood estimation with MCMC sampling.  Using the \texttt{emcee} package developed by  \citet{emcee}, we maximised the log of the likelihood, which is expressed as: 
\begin{gather}
\ln{\mathcal{L}}  = -0.5 \sum_{i,j} \left[ \ln{\sigdij }  + \frac{ \left(\log{T_{i,j}}- \log{T_{{\rm m}, i, j}} \right)^2  } { \sigdij} \right] \\ 
T_{{\rm m}, i, j}  =  T_{\rm m}(\xij) \\
  \sigdij  =  \sigma^2_{\rm int}(\xij)    +  \sigma^2_{\rm stat,i,j}.
\end{gather}
The fit marginalises over a total of fifteen parameters: eight describing the shape of the median profile $(T_{\rm min}, T_{0}, x_t, x_{\rm cool}, a, a_{\rm cool}, b, c)$, and seven additional parameters describing the intrinsic scatter profile. We used flat priors on all parameters.

\begin{table}[!t]
\caption{\footnotesize Parameters and marginalised 68\% uncertainties of the best-fitting analytical model (Eqns. 2-5), and the natural logarithm of the radial intrinsic scatter term, for the fits to the \excpres\ and \rexcess\ samples. The fits are illustrated in Fig.~\ref{f:sct}.}
\begin{center}
\begin{tabular}{lcc}
\toprule

Parameter  &  \gwpfont{EXCPReS}  &  \gwpfont{REXCESS}  \\

\midrule
  $T_{\rm min}$ &   $0.509_{-0.452}^{+0.452}$	&  $0.460_{-0.377}^{+0.377}$ \\
  $T_0$ 		&    $1.161_{-0.099}^{+0.149}$  & $1.160_{-0.101}^{+0.097}$\\
  $x$ 		&    $1.095_{-0.485}^{+0.581}$	&  $1.261_{-0.379}^{+0.391}$\\
  $x_{\rm cool}$ &   $0.039_{-0.254}^{+0.185}$	&  $0.101_{-0.155}^{+0.177}$\\
  $a$ 		&    $0.036_{-0.027}^{+0.070}$	&  $0.043_{-0.033}^{+0.067}$\\
  $a_{\rm cool}$ &   $0.898_{-0.378}^{+0.553}$ &  $1.282_{-0.369}^{+0.514}$\\
  $b$			 &   $2.360_{-0.533}^{+0.992}$ &  $2.318_{-0.449}^{+0.537}$\\
  $c$			 &   $2.219_{-1.145}^{+1.600}$ &  $3.130_{-1.272}^{+1.064}$\\
  \midrule
  $\sigma_1$ 	&  $0.648_{-0.348}^{+0.336}$ &  $0.392_{-0.126}^{+0.184}$\\
 $\sigma_2$	&  $0.374_{-0.081}^{+0.099}$ &  $0.274_{-0.049}^{+0.064}$\\
 $\sigma_3$	&  $0.173_{-0.040}^{+0.051}$ &  $0.139_{-0.021}^{+0.024}$\\
 $\sigma_4$	&  $0.091_{-0.014}^{+0.017}$ &  $0.099_{-0.013}^{+0.014}$\\
 $\sigma_5$	&  $0.060_{-0.012}^{+0.012}$ &  $0.066_{-0.008}^{+0.009}$\\
 $\sigma_6$	&  $0.058_{-0.011}^{+0.011}$ &  $0.031_{-0.012}^{+0.013}$\\
 $\sigma_7$	&  $0.225_{0.033}^{+0.044}$ &   $0.143_{-0.033}^{+0.036}$\\

\bottomrule
\end{tabular}
\end{center}
\label{t:pars}
 \end{table}

The orange envelope in Figure~\ref{f:sct} shows the resulting best-fitting model overplotted on the \excpres\ data points. The best-fitting parameters for $f(x)$ are as follows:
\begin{dmath}
   (T_{\rm min}, T_{0}, x_t, x_{\rm cool}, a, a_{\rm cool}, b, c) = (0.51, 1.16, 1.09, 0.04, 0.04, 0.90, 2.36, 2.22)
\end{dmath}.
The marginalised posterior distributions of each parameter are shown in Fig.~\ref{f:corner}. For comparison, we also show in Fig.~\ref{f:sct} 
 the best-fitting model obtained from fitting the 3D temperature profiles of the representative local X-ray-selected \rexcess\ sample \citep{pra07}. Finally, we also overplot the best-fitting model to the local cool-core sample published by \citet{vik06}. Outside the central region, the agreement between models is good, suggesting no evolution in the bulk temperature profile within the redshift range probed by the present samples. Agreement in the outer regions is expected from theoretical models of self-similarity. However, within the central $\sim 0.2\,R_{500}$ region, there is a suggestion that the cool core sample of \citet{vik06} has a lower central temperature and a higher peak temperature than either of the representative samples. Here, comparison between samples is hampered by radial binning considerations, where cool core systems always have a finer binning than non-cool core systems because of their higher S/N. Further progress on this issue will necessitate careful treatment of the S/N in the central regions, and the inclusion of the dynamical state as an additional parameter \citep[e.g.][]{bar19}.

\section{Summary and conclusions}

In this paper, we introduce \excpres, a representative X-ray selected sample of 31 galaxy clusters at moderate redshifts ($0.4 < z < 0.6$), and spanning the full mass range ($10^{14} < M_{500} < 10^{15}$~M$_{\odot}$). \excpres\ was constructed to be an analogue of the low-redshift \rexcess\ sample \citep{boe07}. 

We used the \xmm\ observations of the \excpres\ sample to develop and test a new method to produce optimally binned X-ray temperature profiles. The method uses a dynamic programming algorithm based on partitioning of the $S_{\rm X}$-bin range to obtain a new binning scheme that fulfils a given S/N threshold criterion out to large radius. Additional optional criteria can be included, including logarithmic radial binning, or setting a minimum number of $S_{\rm X}$-bins to be included in each $kT$-bin. The method aims at maximising the number of temperature profile bins out to the largest (optionally required) radius. A user-chosen minimum number of bins can be used as a fallback solution for cases with low S/N: in this case, the algorithm maximises the S/N of all bins simultaneously.

We demonstrated the efficiency of our method using the \excpres\ sample, which contains data sets covering a wide range of S/N. The expected correlation between the S/N in $kT$-bins and the relative error on the temperature shows very little dispersion, but exhibits a clear trend with global temperature. Combining the results from \excpres\ and \rexcess, we derived a relation between the relative uncertainty in $kT$-bins with respect to the soft-band $S_{\rm X}$ S/N within these $kT$-bins, and its dependence on global temperature and redshift (Eq.~\ref{e:dtt}). This relation provides a useful tool for exposure time calculation in X-ray observations, allowing one to obtain an estimate of a given relative error on the ICM temperature measurements, based only on the knowledge of the soft-band number counts.

The optimally binned 2D temperature profiles were PSF-deconvolved and deprojected to derive the 3D profiles. Once scaled by $R_{500}$ and $T_{\rm X}$, the temperature in the $[0.15-0.75]\,R_{500}$ region, the 3D \excpres\ temperature profiles exhibit a clear self similar behaviour beyond the core region and increased dispersion towards the centre. We obtained a mean temperature profile for the \excpres\ sample and compared to that from the local X-ray selected \rexcess\ sample. This comparison shows no obvious sign of evolution in the average temperature profile shape in the redshift range probed in the present study.

In a forthcoming paper we will further investigate the 3D thermodynamic profiles of the \excpres\ sample and how the global properties scale,  with respect to the expected self similar evolution and to the cluster mass.

\begin{acknowledgements}
Santa, xeus (our local computing facility) for crashing only once during all these years...
We would like to thank the people and funding agencies who have supported us over the many years that it has taken us to write this paper. We acknowledge in particular long-term support from CNRS/INSU and from the French Centre National d'\'Etudes Spatiales (CNES). The results reported in this article are based on data obtained from the XMM-{\it Newton} observatory, an ESA science mission with instruments and contributions directly funded by ESA Member States and NASA.

\end{acknowledgements}

\bibliographystyle{aa}
\bibliography{gwpbib2}

\begin{thebibliography}{67}
\expandafter\ifx\csname natexlab\endcsname\relax\def\natexlab#1{#1}\fi

\bibitem[{{Anokhin}(2008)}]{ano08}
{Anokhin}, S.~G. 2008, Advances in Space Research, 42, 576

\bibitem[{{Arnaud}(2008)}]{arn08}
{Arnaud}, M. 2008, in The X-ray Universe 2008, 191

\bibitem[{{Arnaud} {et~al.}(2001){Arnaud}, {Neumann}, {Aghanim}, {Gastaud},
  {Majerowicz}, \& {Hughes}}]{arn01}
{Arnaud}, M., {Neumann}, D.~M., {Aghanim}, N., {et~al.} 2001, \aap, 365, L80

\bibitem[{{Arnaud} {et~al.}(2007){Arnaud}, {Pointecouteau}, \& {Pratt}}]{arn07}
{Arnaud}, M., {Pointecouteau}, E., \& {Pratt}, G.~W. 2007, \aap, 474, L37

\bibitem[{{Arnaud} {et~al.}(2010){Arnaud}, {Pratt}, {Piffaretti},
  {B{\"o}hringer}, {Croston}, \& {Pointecouteau}}]{arn10}
{Arnaud}, M., {Pratt}, G.~W., {Piffaretti}, R., {et~al.} 2010, \aap, 517, A92

\bibitem[{Art \& Mauch(2007)}]{art07}
Art, L. \& Mauch, H. 2007, Dynamic Programming: A Computational Tool, Studies
  in Computational Intelligence (Springer Berlin, Heidelberg)

\bibitem[{{Baldi} {et~al.}(2007){Baldi}, {Ettori}, {Mazzotta}, {Tozzi}, \&
  {Borgani}}]{bal07}
{Baldi}, A., {Ettori}, S., {Mazzotta}, P., {Tozzi}, P., \& {Borgani}, S. 2007,
  \apj, 666, 835

\bibitem[{{Baldi} {et~al.}(2012){Baldi}, {Ettori}, {Molendi}, \&
  {Gastaldello}}]{bal12}
{Baldi}, A., {Ettori}, S., {Molendi}, S., \& {Gastaldello}, F. 2012, \aap, 545,
  A41

\bibitem[{{Bartalucci} {et~al.}(2019){Bartalucci}, {Arnaud}, {Pratt},
  {D{\'e}mocl{\`e}s}, \& {Lovisari}}]{bar19}
{Bartalucci}, I., {Arnaud}, M., {Pratt}, G.~W., {D{\'e}mocl{\`e}s}, J., \&
  {Lovisari}, L. 2019, \aap, 628, A86

\bibitem[{{Bartalucci} {et~al.}(2017){Bartalucci}, {Arnaud}, {Pratt},
  {D{\'e}mocl{\`e}s}, {van der Burg}, \& {Mazzotta}}]{bar17}
{Bartalucci}, I., {Arnaud}, M., {Pratt}, G.~W., {et~al.} 2017, \aap, 598, A61

\bibitem[{{Bartalucci} {et~al.}(2018){Bartalucci}, {Arnaud}, {Pratt}, \& {Le
  Brun}}]{bar18}
{Bartalucci}, I., {Arnaud}, M., {Pratt}, G.~W., \& {Le Brun}, A.~M.~C. 2018,
  \aap, 617, A64

\bibitem[{{Bleem} {et~al.}(2015){Bleem}, {Stalder}, {de Haan}, {Aird}, {Allen},
  {Applegate}, {Ashby}, {Bautz}, {Bayliss}, {Benson}, {Bocquet}, {Brodwin},
  {Carlstrom}, {Chang}, {Chiu}, {Cho}, {Clocchiatti}, {Crawford}, {Crites},
  {Desai}, {Dietrich}, {Dobbs}, {Foley}, {Forman}, {George}, {Gladders},
  {Gonzalez}, {Halverson}, {Hennig}, {Hoekstra}, {Holder}, {Holzapfel},
  {Hrubes}, {Jones}, {Keisler}, {Knox}, {Lee}, {Leitch}, {Liu}, {Lueker},
  {Luong-Van}, {Mantz}, {Marrone}, {McDonald}, {McMahon}, {Meyer}, {Mocanu},
  {Mohr}, {Murray}, {Padin}, {Pryke}, {Reichardt}, {Rest}, {Ruel}, {Ruhl},
  {Saliwanchik}, {Saro}, {Sayre}, {Schaffer}, {Schrabback}, {Shirokoff},
  {Song}, {Spieler}, {Stanford}, {Staniszewski}, {Stark}, {Story}, {Stubbs},
  {Vand erlinde}, {Vieira}, {Vikhlinin}, {Williamson}, {Zahn}, \&
  {Zenteno}}]{ble15}
{Bleem}, L.~E., {Stalder}, B., {de Haan}, T., {et~al.} 2015, \apjs, 216, 27

\bibitem[{{B{\"o}hringer} {et~al.}(2004){B{\"o}hringer}, {Schuecker}, {Guzzo},
  {Collins}, {Voges}, {Cruddace}, {Ortiz-Gil}, {Chincarini}, {De Grandi},
  {Edge}, {MacGillivray}, {Neumann}, {Schindler}, \& {Shaver}}]{boh04}
{B{\"o}hringer}, H., {Schuecker}, P., {Guzzo}, L., {et~al.} 2004, \aap, 425,
  367

\bibitem[{{B{\"o}hringer} {et~al.}(2007){B{\"o}hringer}, {Schuecker}, {Pratt},
  {Arnaud}, {Ponman}, {Croston}, {Borgani}, {Bower}, {Briel}, {Collins},
  {Donahue}, {Forman}, {Finoguenov}, {Geller}, {Guzzo}, {Henry}, {Kneissl},
  {Mohr}, {Matsushita}, {Mullis}, {Ohashi}, {Pedersen}, {Pierini}, {Quintana},
  {Raychaudhury}, {Reiprich}, {Romer}, {Rosati}, {Sabirli}, {Temple}, {Viana},
  {Vikhlinin}, {Voit}, \& {Zhang}}]{boe07}
{B{\"o}hringer}, H., {Schuecker}, P., {Pratt}, G.~W., {et~al.} 2007, \aap, 469,
  363

\bibitem[{{B{\"o}hringer} {et~al.}(2000){B{\"o}hringer}, {Voges}, {Huchra},
  {McLean}, {Giacconi}, {Rosati}, {Burg}, {Mader}, {Schuecker}, {Simi{\c c}},
  {Komossa}, {Reiprich}, {Retzlaff}, \& {Tr{\"u}mper}}]{boh00}
{B{\"o}hringer}, H., {Voges}, W., {Huchra}, J.~P., {et~al.} 2000, \apjs, 129,
  435

\bibitem[{{Burenin} {et~al.}(2007){Burenin}, {Vikhlinin}, {Hornstrup},
  {Ebeling}, {Quintana}, \& {Mescheryakov}}]{bur07}
{Burenin}, R.~A., {Vikhlinin}, A., {Hornstrup}, A., {et~al.} 2007, \apjs, 172,
  561

\bibitem[{{Burke} {et~al.}(2003){Burke}, {Collins}, {Sharples}, {Romer}, \&
  {Nichol}}]{bur03}
{Burke}, D.~J., {Collins}, C.~A., {Sharples}, R.~M., {Romer}, A.~K., \&
  {Nichol}, R.~C. 2003, \mnras, 341, 1093

\bibitem[{{Cormen} {et~al.}(2009){Cormen}, {Leiserson}, {Rivest}, \&
  {Stein}}]{clr}
{Cormen}, T.~H., {Leiserson}, C.~E., {Rivest}, R.~L., \& {Stein}, C. 2009,
  {Introduction to Algorithms} (MIT Press and McGraw-Hill)

\bibitem[{{Croston} {et~al.}(2006){Croston}, {Arnaud}, {Pointecouteau}, \&
  {Pratt}}]{cro06}
{Croston}, J.~H., {Arnaud}, M., {Pointecouteau}, E., \& {Pratt}, G.~W. 2006,
  \aap, 459, 1007

\bibitem[{{De Grandi} \& {Molendi}(2002)}]{deg02}
{De Grandi}, S. \& {Molendi}, S. 2002, \apj, 567, 163

\bibitem[{{Ebeling} {et~al.}(2007){Ebeling}, {Barrett}, {Donovan}, {Ma},
  {Edge}, \& {van Speybroeck}}]{ebe07}
{Ebeling}, H., {Barrett}, E., {Donovan}, D., {et~al.} 2007, \apjl, 661, L33

\bibitem[{{Ebeling} {et~al.}(2010){Ebeling}, {Edge}, {Mantz}, {Barrett},
  {Henry}, {Ma}, \& {van Speybroeck}}]{ebe10}
{Ebeling}, H., {Edge}, A.~C., {Mantz}, A., {et~al.} 2010, \mnras, 407, 83

\bibitem[{{Foreman-Mackey} {et~al.}(2013){Foreman-Mackey}, {Hogg}, {Lang}, \&
  {Goodman}}]{emcee}
{Foreman-Mackey}, D., {Hogg}, D.~W., {Lang}, D., \& {Goodman}, J. 2013, \pasp,
  125, 306

\bibitem[{{Gioia} {et~al.}(1990){Gioia}, {Maccacaro}, {Schild}, {Wolter},
  {Stocke}, {Morris}, \& {Henry}}]{gio90}
{Gioia}, I.~M., {Maccacaro}, T., {Schild}, R.~E., {et~al.} 1990, \apjs, 72, 567

\bibitem[{{Hasselfield} {et~al.}(2013){Hasselfield}, {Hilton}, {Marriage},
  {Addison}, {Barrientos}, {Battaglia}, {Battistelli}, {Bond}, {Crichton},
  {Das}, {Devlin}, {Dicker}, {Dunkley}, {D{\"u}nner}, {Fowler}, {Gralla},
  {Hajian}, {Halpern}, {Hincks}, {Hlozek}, {Hughes}, {Infante}, {Irwin},
  {Kosowsky}, {Marsden}, {Menanteau}, {Moodley}, {Niemack}, {Nolta}, {Page},
  {Partridge}, {Reese}, {Schmitt}, {Sehgal}, {Sherwin}, {Sievers}, {Sif{\'o}n},
  {Spergel}, {Staggs}, {Swetz}, {Switzer}, {Thornton}, {Trac}, \&
  {Wollack}}]{has13}
{Hasselfield}, M., {Hilton}, M., {Marriage}, T.~A., {et~al.} 2013, \jcap, 2013,
  008

\bibitem[{{Henry}(2004)}]{hen04}
{Henry}, J.~P. 2004, \apj, 609, 603

\bibitem[{{Henry} {et~al.}(2006){Henry}, {Mullis}, {Voges}, {B{\"o}hringer},
  {Briel}, {Gioia}, \& {Huchra}}]{hen06}
{Henry}, J.~P., {Mullis}, C.~R., {Voges}, W., {et~al.} 2006, \apjs, 162, 304

\bibitem[{{Horner} {et~al.}(2008){Horner}, {Perlman}, {Ebeling}, {Jones},
  {Scharf}, {Wegner}, {Malkan}, \& {Maughan}}]{hor08}
{Horner}, D.~J., {Perlman}, E.~S., {Ebeling}, H., {et~al.} 2008, \apjs, 176,
  374

\bibitem[{{Kalberla} {et~al.}(2005){Kalberla}, {Burton}, {Hartmann}, {Arnal},
  {Bajaja}, {Morras}, \& {P{\"o}ppel}}]{lab}
{Kalberla}, P.~M.~W., {Burton}, W.~B., {Hartmann}, D., {et~al.} 2005, \aap,
  440, 775

\bibitem[{{Kay} \& {Pratt}(2022)}]{kay22}
{Kay}, S.~T. \& {Pratt}, G.~W. 2022, in Handbook of X-ray and Gamma-ray
  Astrophysics. Edited by Cosimo Bambi and Andrea Santangelo, 100

\bibitem[{{Kotov} \& {Vikhlinin}(2005)}]{kot05}
{Kotov}, O. \& {Vikhlinin}, A. 2005, \apj, 633, 781

\bibitem[{{Kravtsov} \& {Borgani}(2012)}]{kra12}
{Kravtsov}, A.~V. \& {Borgani}, S. 2012, \araa, 50, 353

\bibitem[{{Kravtsov} {et~al.}(2006){Kravtsov}, {Vikhlinin}, \& {Nagai}}]{kra06}
{Kravtsov}, A.~V., {Vikhlinin}, A., \& {Nagai}, D. 2006, \apj, 650, 128

\bibitem[{{Leccardi} \& {Molendi}(2008)}]{lec08}
{Leccardi}, A. \& {Molendi}, S. 2008, \aap, 486, 359

\bibitem[{{Lovisari} \& {Maughan}(2022)}]{lov22}
{Lovisari}, L. \& {Maughan}, B.~J. 2022, in Handbook of X-ray and Gamma-ray
  Astrophysics. Edited by Cosimo Bambi and Andrea Santangelo, 65

\bibitem[{{Lovisari} {et~al.}(2015){Lovisari}, {Reiprich}, \&
  {Schellenberger}}]{lov15}
{Lovisari}, L., {Reiprich}, T.~H., \& {Schellenberger}, G. 2015, \aap, 573,
  A118

\bibitem[{{Lumb} {et~al.}(2004){Lumb}, {Bartlett}, {Romer}, {Blanchard},
  {Burke}, {Collins}, {Nichol}, {Giard}, {Marty}, {Nevalainen}, {Sadat}, \&
  {Vauclair}}]{lum04}
{Lumb}, D.~H., {Bartlett}, J.~G., {Romer}, A.~K., {et~al.} 2004, \aap, 420, 853

\bibitem[{{Mann} \& {Ebeling}(2012)}]{man12}
{Mann}, A.~W. \& {Ebeling}, H. 2012, \mnras, 420, 2120

\bibitem[{{Mantz} {et~al.}(2016){Mantz}, {Allen}, {Morris}, \&
  {Schmidt}}]{man16}
{Mantz}, A.~B., {Allen}, S.~W., {Morris}, R.~G., \& {Schmidt}, R.~W. 2016,
  \mnras, 456, 4020

\bibitem[{{Markevitch}(1998)}]{mar98}
{Markevitch}, M. 1998, \apj, 504, 27

\bibitem[{{McDonald} {et~al.}(2014){McDonald}, {Benson}, {Vikhlinin}, {Aird},
  {Allen}, {Bautz}, {Bayliss}, {Bleem}, {Bocquet}, {Brodwin}, {Carlstrom},
  {Chang}, {Cho}, {Clocchiatti}, {Crawford}, {Crites}, {de Haan}, {Dobbs},
  {Foley}, {Forman}, {George}, {Gladders}, {Gonzalez}, {Halverson},
  {Hlavacek-Larrondo}, {Holder}, {Holzapfel}, {Hrubes}, {Jones}, {Keisler},
  {Knox}, {Lee}, {Leitch}, {Liu}, {Lueker}, {Luong-Van}, {Mantz}, {Marrone},
  {McMahon}, {Meyer}, {Miller}, {Mocanu}, {Mohr}, {Murray}, {Padin}, {Pryke},
  {Reichardt}, {Rest}, {Ruhl}, {Saliwanchik}, {Saro}, {Sayre}, {Schaffer},
  {Shirokoff}, {Spieler}, {Stalder}, {Stanford}, {Staniszewski}, {Stark},
  {Story}, {Stubbs}, {Vanderlinde}, {Vieira}, {Williamson}, {Zahn}, \&
  {Zenteno}}]{mac14}
{McDonald}, M., {Benson}, B.~A., {Vikhlinin}, A., {et~al.} 2014, \apj, 794, 67

\bibitem[{{Melin} {et~al.}(2006){Melin}, {Bartlett}, \& {Delabrouille}}]{mel06}
{Melin}, J.~B., {Bartlett}, J.~G., \& {Delabrouille}, J. 2006, \aap, 459, 341

\bibitem[{{Mullis} {et~al.}(2003){Mullis}, {McNamara}, {Quintana}, {Vikhlinin},
  {Henry}, {Gioia}, {Hornstrup}, {Forman}, \& {Jones}}]{mul03}
{Mullis}, C.~R., {McNamara}, B.~R., {Quintana}, H., {et~al.} 2003, \apj, 594,
  154

\bibitem[{{Perlman} {et~al.}(2002){Perlman}, {Horner}, {Jones}, {Scharf},
  {Ebeling}, {Wegner}, \& {Malkan}}]{per02}
{Perlman}, E.~S., {Horner}, D.~J., {Jones}, L.~R., {et~al.} 2002, \apjs, 140,
  265

\bibitem[{{Piffaretti} {et~al.}(2011){Piffaretti}, {Arnaud}, {Pratt},
  {Pointecouteau}, \& {Melin}}]{pif11}
{Piffaretti}, R., {Arnaud}, M., {Pratt}, G.~W., {Pointecouteau}, E., \&
  {Melin}, J.-B. 2011, \aap, 534, A109

\bibitem[{{Planck Collaboration VIII}(2011)}]{esz}
{Planck Collaboration VIII}. 2011, \aap, 536, A8

\bibitem[{{Planck Collaboration XXIX}(2014)}]{psz1}
{Planck Collaboration XXIX}. 2014, \aap, 571, A29

\bibitem[{{Planck Collaboration XXVII}(2016)}]{PSZ2}
{Planck Collaboration XXVII}. 2016, \aap, 594, A27

\bibitem[{{Pratt} \& {Arnaud}(2003)}]{pra03}
{Pratt}, G.~W. \& {Arnaud}, M. 2003, \aap, 408, 1

\bibitem[{{Pratt} {et~al.}(2022){Pratt}, {Arnaud}, {Maughan}, \&
  {Melin}}]{pra22}
{Pratt}, G.~W., {Arnaud}, M., {Maughan}, B.~J., \& {Melin}, J.~B. 2022, \aap,
  665, A24

\bibitem[{{Pratt} {et~al.}(2010){Pratt}, {Arnaud}, {Piffaretti},
  {B{\"o}hringer}, {Ponman}, {Croston}, {Voit}, {Borgani}, \& {Bower}}]{pra10}
{Pratt}, G.~W., {Arnaud}, M., {Piffaretti}, R., {et~al.} 2010, \aap, 511, A85+

\bibitem[{{Pratt} {et~al.}(2007){Pratt}, {B{\"o}hringer}, {Croston}, {Arnaud},
  {Borgani}, {Finoguenov}, \& {Temple}}]{pra07}
{Pratt}, G.~W., {B{\"o}hringer}, H., {Croston}, J.~H., {et~al.} 2007, \aap,
  461, 71

\bibitem[{{Repp} \& {Ebeling}(2018)}]{rep18}
{Repp}, A. \& {Ebeling}, H. 2018, \mnras, 479, 844

\bibitem[{{Romer} {et~al.}(2000){Romer}, {Nichol}, {Holden}, {Ulmer}, {Pildis},
  {Merrelli}, {Adami}, {Burke}, {Collins}, {Metevier}, {Kron}, \&
  {Commons}}]{rom00}
{Romer}, A.~K., {Nichol}, R.~C., {Holden}, B.~P., {et~al.} 2000, \apjs, 126,
  209

\bibitem[{{Rykoff} {et~al.}(2016){Rykoff}, {Rozo}, {Hollowood},
  {Bermeo-Hernandez}, {Jeltema}, {Mayers}, {Romer}, {Rooney}, {Saro}, {Vergara
  Cervantes}, {Wechsler}, {Wilcox}, {Abbott}, {Abdalla}, {Allam}, {Annis},
  {Benoit-L{\'e}vy}, {Bernstein}, {Bertin}, {Brooks}, {Burke}, {Capozzi},
  {Carnero Rosell}, {Carrasco Kind}, {Castander}, {Childress}, {Collins},
  {Cunha}, {D'Andrea}, {da Costa}, {Davis}, {Desai}, {Diehl}, {Dietrich},
  {Doel}, {Evrard}, {Finley}, {Flaugher}, {Fosalba}, {Frieman}, {Glazebrook},
  {Goldstein}, {Gruen}, {Gruendl}, {Gutierrez}, {Hilton}, {Honscheid}, {Hoyle},
  {James}, {Kay}, {Kuehn}, {Kuropatkin}, {Lahav}, {Lewis}, {Lidman}, {Lima},
  {Maia}, {Mann}, {Marshall}, {Martini}, {Melchior}, {Miller}, {Miquel},
  {Mohr}, {Nichol}, {Nord}, {Ogando}, {Plazas}, {Reil}, {Sahl{\'e}n},
  {Sanchez}, {Santiago}, {Scarpine}, {Schubnell}, {Sevilla-Noarbe}, {Smith},
  {Soares-Santos}, {Sobreira}, {Stott}, {Suchyta}, {Swanson}, {Tarle},
  {Thomas}, {Tucker}, {Uddin}, {Viana}, {Vikram}, {Walker}, {Zhang}, \& {DES
  Collaboration}}]{ryk16}
{Rykoff}, E.~S., {Rozo}, E., {Hollowood}, D., {et~al.} 2016, \apjs, 224, 1

\bibitem[{{Sadibekova} {et~al.}(2023){Sadibekova}, {Arnaud}, {Pratt}, {Melin},
  \& {P. Tarr\'io}}]{sad23}
{Sadibekova}, T., {Arnaud}, M., {Pratt}, G.~W., {Melin}, J., \& {P. Tarr\'io},
  P. 2023, \aap

\bibitem[{{Sadibekova} {et~al.}(2024){Sadibekova}, {Arnaud}, {Pratt},
  {Tarr{\'\i}o}, \& {Melin}}]{sad24}
{Sadibekova}, T., {Arnaud}, M., {Pratt}, G.~W., {Tarr{\'\i}o}, P., \& {Melin},
  J.~B. 2024, arXiv e-prints, arXiv:2402.01538

\bibitem[{{Schaye} {et~al.}(2015){Schaye}, {Crain}, {Bower}, {Furlong},
  {Schaller}, {Theuns}, {Dalla Vecchia}, {Frenk}, {McCarthy}, {Helly},
  {Jenkins}, {Rosas-Guevara}, {White}, {Baes}, {Booth}, {Camps}, {Navarro},
  {Qu}, {Rahmati}, {Sawala}, {Thomas}, \& {Trayford}}]{sch15}
{Schaye}, J., {Crain}, R.~A., {Bower}, R.~G., {et~al.} 2015, \mnras, 446, 521

\bibitem[{{Schaye} {et~al.}(2023){Schaye}, {Kugel}, {Schaller}, {Helly},
  {Braspenning}, {Elbers}, {McCarthy}, {van Daalen}, {Vandenbroucke}, {Frenk},
  {Kwan}, {Salcido}, {Bah{\'e}}, {Borrow}, {Chaikin}, {Hahn}, {Hu{\v{s}}ko},
  {Jenkins}, {Lacey}, \& {Nobels}}]{sch23}
{Schaye}, J., {Kugel}, R., {Schaller}, M., {et~al.} 2023, arXiv e-prints,
  arXiv:2306.04024

\bibitem[{{Sun} {et~al.}(2009){Sun}, {Voit}, {Donahue}, {Jones}, {Forman}, \&
  {Vikhlinin}}]{sun09}
{Sun}, M., {Voit}, G.~M., {Donahue}, M., {et~al.} 2009, \apj, 693, 1142

\bibitem[{{Ulmer} {et~al.}(2005){Ulmer}, {Adami}, {Covone}, {Durret}, {Lima
  Neto}, {Sabirli}, {Holden}, {Kron}, \& {Romer}}]{ulm05}
{Ulmer}, M.~P., {Adami}, C., {Covone}, G., {et~al.} 2005, \apj, 624, 124

\bibitem[{{Vikhlinin}(2006)}]{vik06b}
{Vikhlinin}, A. 2006, \apj, 640, 710

\bibitem[{{Vikhlinin} {et~al.}(2006){Vikhlinin}, {Kravtsov}, {Forman}, {Jones},
  {Markevitch}, {Murray}, \& {Van Speybroeck}}]{vik06}
{Vikhlinin}, A., {Kravtsov}, A., {Forman}, W., {et~al.} 2006, \apj, 640, 691

\bibitem[{{Vikhlinin} {et~al.}(2005){Vikhlinin}, {Markevitch}, {Murray},
  {Jones}, {Forman}, \& {Van Speybroeck}}]{vik05}
{Vikhlinin}, A., {Markevitch}, M., {Murray}, S.~S., {et~al.} 2005, \apj, 628,
  655

\bibitem[{{Vikhlinin} {et~al.}(1998){Vikhlinin}, {McNamara}, {Forman}, {Jones},
  {Quintana}, \& {Hornstrup}}]{vik98}
{Vikhlinin}, A., {McNamara}, B.~R., {Forman}, W., {et~al.} 1998, \apj, 502, 558

\bibitem[{{Vogelsberger} {et~al.}(2014){Vogelsberger}, {Genel}, {Springel},
  {Torrey}, {Sijacki}, {Xu}, {Snyder}, {Nelson}, \& {Hernquist}}]{vog14}
{Vogelsberger}, M., {Genel}, S., {Springel}, V., {et~al.} 2014, \mnras, 444,
  1518

\bibitem[{{Wen} \& {Han}(2015)}]{wen15}
{Wen}, Z.~L. \& {Han}, J.~L. 2015, \apj, 807, 178

\end{thebibliography}

\begin{appendix}

\section{Clusters with \xmm\ observations not included in \excpres}
\label{ap:discard}

\begin{table*}[t]
\caption{\footnotesize Clusters in the $0.4<z<0.6$ range with \xmm\ archival data but not included in the \excpres\ sample. The clusters are ordered by increasing redshift. Columns are 1-2: Cluster MCXC name and other name, 3: Detection survey, 4: Redshift, 5: \xmm\ OBSID, 6: Reference to relevant publication or present work; 
(1) \citet{bal12};  (2) \citet{ano08};  (3) \citet{lum04}; and (4) \citet{ulm05}} 
\begin{center}
\resizebox{0.95\textwidth}{!} {
\begin{tabular}{lllccl}
\toprule
 Name & Other Name & Catalogue & z &  OBSID & Ref \\ 
\midrule
   MCXC J0305.3+1728	& MS0302.5+1717		&	EMSS   					&    0.425 &  0112190101 & 1         \\ 
   MCXC J2202.7-1902 	&	RX J2202.7-1902	&	160SD, S-SHARC, WARPSII 	&    0.438 &  0203450201 & 2, pw    \\
   MCXC J1325.5-3825 	&	RX J1325.5-3826 	&	S-SHARC   				&    0.445 &  0110890101  	& 3 \\
   MCXC J0858.4+1357 	&	RX J0858.4+1357	 &	160SD, S-SHARC, WARPSII 	&    0.488 &  0203450101 &  2, pw	\\
   MCXC J0505.3-2849 	&	RX J0505.3-2849 	&	S-SHARC   			   	&    0.509 &  0111160201 &	3, pw \\
   MCXC J1002.6-0808 	&	RX J1002.6-0808 	&	160SD, WARPSII			 &    0.524 &  0302580301 & pw	 \\
   MCXC J1354.2-0221 	&	BVH2007 181 		&	400SD, S-SHARC, 160SD	&    0.546 &  0112250101 &	3\\
   MCXC J0847.1+3449	 &	RX J0847.1+3449 	&	160SD					&    0.560 &  0107860501 &  3 \\
   MCXC J0056.9-2740 	&	RX J0056.9-2740 	&	160SD					&    0.563 &  0111280201 &  pw \\   
            &    	&      &      		&   0111282001 &       \\ 
   MCXC J1419.8+0634 	&   WARP J1419.9+0634 &	WARPSII, 160SD 			&    0.564 &  0303670101 &  1, pw \\
   MCXC J2056.3-0437 	&        MS2053.7-0449 	&	 EMSS   					&    0.583 &  0112190601 &  1 \\
   MCXC J0337.7-2522 	&      RX J0337.7-2522 	&              160SD, S-SHARC   		&    0.585 &  0107860401 &  3  \\
   MCXC J1205.8+4429 	&    WARP J1205.8+4429 &                      WARPSII 		 	&    0.592 &  0156360101 & 	 4 \\ 
 \bottomrule
\end{tabular}
 }
\end{center}
\label{tab:discard}
 \end{table*}

The  MCXC clusters in the  low-luminosity sample ($\Lv < 4\times 10^{44}$~ergs~s$^{-1}$) with archival observations, but not included  in the \excpres\ sample,  are listed in Tab.~\ref{tab:discard}. The table also gives the OBSID of the \xmm\ observation and the reference of  relevant published   \xmm\ analysis. We further analysed the archival data of some of the clusters.

MCXC J1002.6-0808 was observed in the framework of our Large Programme \#$030258$. The observation revealed that MCXC J1002.6-0808 is a point source, which is also confirmed by the \chandra\ image.

Our study of the image of MCXC J1419.8+0634 shows that it is a bimodal cluster, thus not suitable for radial analysis.  

We require a minimal S/N ratio of $S/N=20$,  needed to derive a  temperature profile as shown by our work. The other clusters could not be  retained because the archival  observations  are not deep enough:
 \begin{itemize}      
 \item  The observations of the two EMSS clusters,  MCXC J0305.3+1728 ($z=0.425$) and MCXC J2056.3-0437 ($z=0.583$, are shallow, with clean observing time of $\sim 10$ ksec and $12$\,ksec, respectively \citep[][their Tab.1]{bal12}. The error on the global temperature is  $\pm15\%$.  \citet{bal12} were only able to extract the temperature in 2 bins up to $0.4\RV$ for MCXC J2056.3-0437.

\item  The observations of MCXC J1325.5-3825,  MCXC J0505.3-2849, MCXC J1354.2-0221, MCXC J0847.1+3449 and  MCXC J0337.7-2522 were early follow up of SHARC clusters to meant to measure a global temperature and the luminosity,  published by \citet{lum04}. Our full re-analysis of  MCXC J0505.3-2849, the best measured cluster of the list, gives a S/N$_{500}=20$. The observations of the other clusters are at lower S/N, taking into account the precision on the observed flux (their table 5).  MCXC J1325.5-3825 falls in the field of view of an observation centered on IRAS 13224$-$3809,  a very bright Seyfert galaxy. Many more observations have become available over the years, but only in Window mode,  and centred  on the galaxy.

Note that the two other clusters of their sample at $0.4<z<0.6$, MCXC J1120.1+4318 and  MCXC J1701.3+6414 (re-observed by \xmm), are  included in the \excpres\ sample. 

\item Together with MCXC J0337.7-2522 studied by \citet{lum04}, MCXC J2202.7-1902 and MCXC J0858.4+1357 are the three $0.4<z<0.6$ clusters  with the lowest luminosities,  $\Lv  < 10^{44}$ergs/s, observed by \xmm.   \citet{ano08} gives  a temperature of about $3$\,keV for MCXC J2202.7-1902 and MCXC J0858.4+1357. From our analysis of the surface brightness profiles and global properties, we derived a S/N ratio of S/N=15 and S/N$_{500}$=19, respectively. This entails that the \excpres\ selection could not be extended  below $\Lv=10^{44}$ergs/s.

\item MCXC J0056.9-2740 ($z=0.56$)  falls in the field of  the programme `A shallow XMM survey of AAT 2DF fields SSC\_32' (PI M. Watson). This includes two short observations of 7.2 ksec (highly flared)  and 8.9 ksec, respectively. This is too shallow to obtain spatially resolved spectroscopy of the $z=0.56$ cluster and indeed the examination  of  the EPIC image shows that the cluster is poorly  detected. MCXC J0056.9-2740 coincides with the source 4XMM J005657.1-274028, detected  at  $13\,\sigma$ in  the 4XMM-DR13 catalogue. 

\item MCXC J1205.8+4429 is a fossil group, as shown by \citet{ulm05}.  From their Table~2, the S/N of the group observation is $S/N \sim 21$ and there is  a $\pm 10\%$ error on the global temperature. 

\end{itemize}

In conclusion, the final selection for the low luminosity box of \excpres\ includes clusters with $0.44<z<0.56$  and $10^{44}<\Lv<4\times10^{44}$ ergs~s$^{-1}$. Four further objects with archival data fall in that range but are not retained: three SHARC clusters published by \citet{lum04} since the exposure is not deep enough, and a 160SD/WARPS object which is false detection.

\section{Individual temperature profiles}
\label{a:indt}

Individual temperature profiles of the \excpres\ sample are shown in Fig.~\ref{f:kTprof1} and ~\ref{f:kTprof2}. In all panels, the black points with error bars show the annular temperature measurements. Solid green lines show the best fitting 3D temperature model, with its uncertainty indicated by the dashed green lines. Red lines show the 2D reprojection of the best fitting 3D model, and dashed red lines the associated uncertainty.

\begin{figure*}[!t]
\hfill
\includegraphics[width=0.975\textwidth]{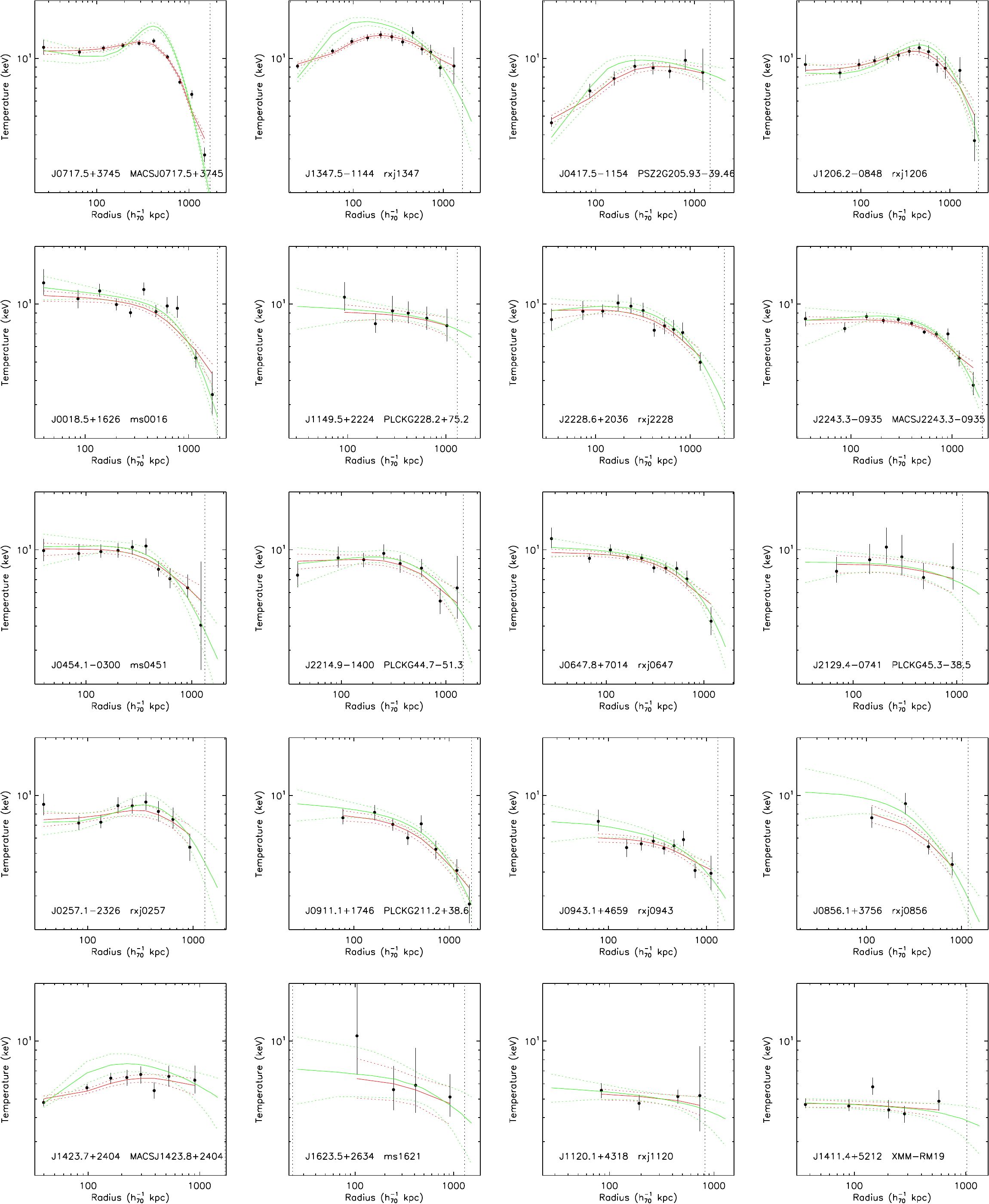}
\caption{\footnotesize  Individual temperature profiles of the \excpres\ sample. Black points with error bars show the annular temperature measurements. Green lines show the best fitting 3D temperature model. Red lines show the 2D reprojection of the best fitting 3D model. The model uncertainties are indicated by dashed lines.}
\label{f:kTprof1}
\end{figure*}

\begin{figure*}[!t]
\hfill
\includegraphics[width=0.975\textwidth]{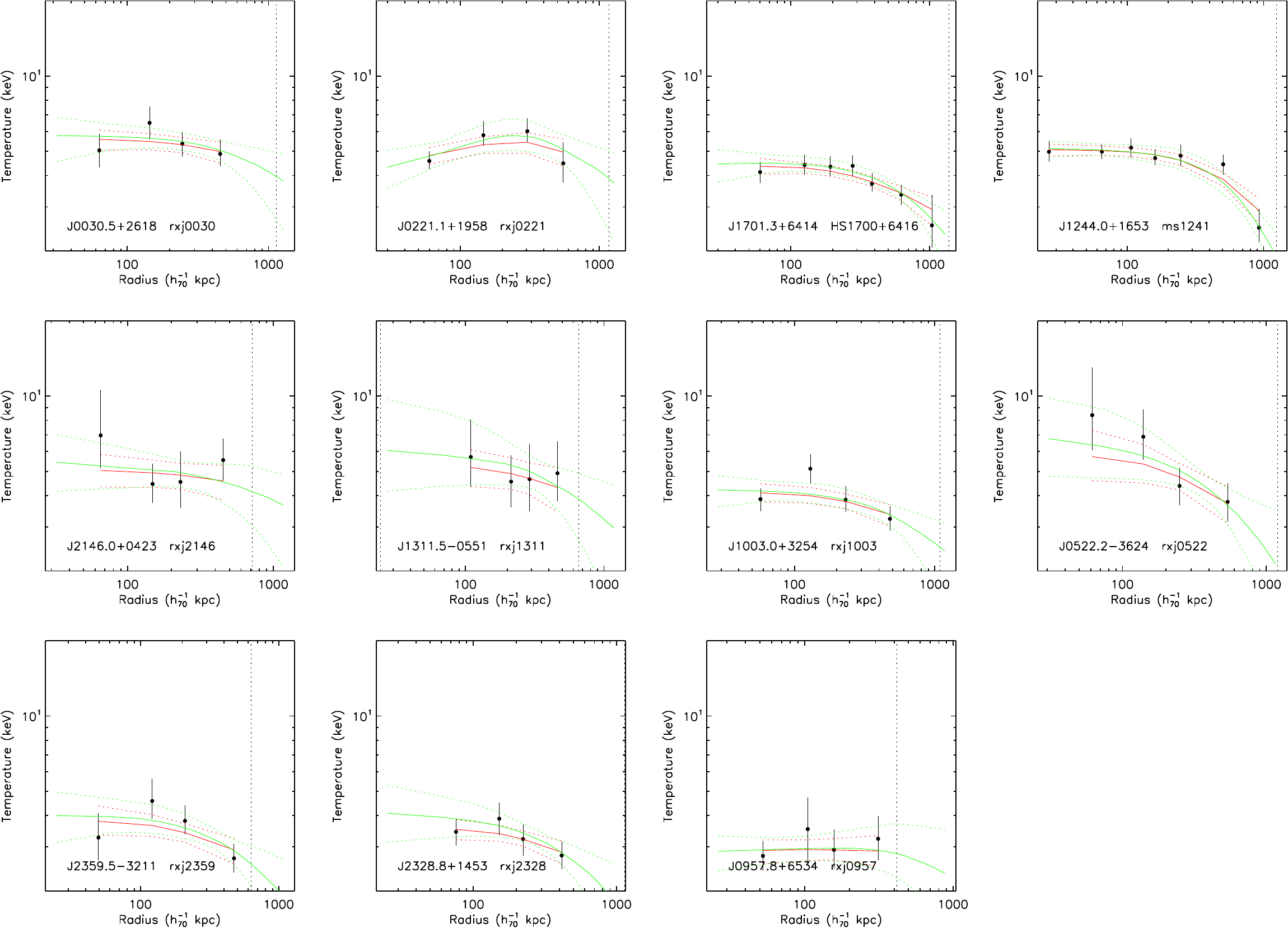}
\caption{\footnotesize continued  }
\label{f:kTprof2}
\end{figure*}

\section{Posteriors}

Marginalised posterior likelihood for the parameters of the best-fitting temperature profile model to the \excpres\ data, as detailed in Sect.~\ref{s:model}.

\begin{figure*}[!t]
\hfill
\includegraphics[width=0.975\textwidth]{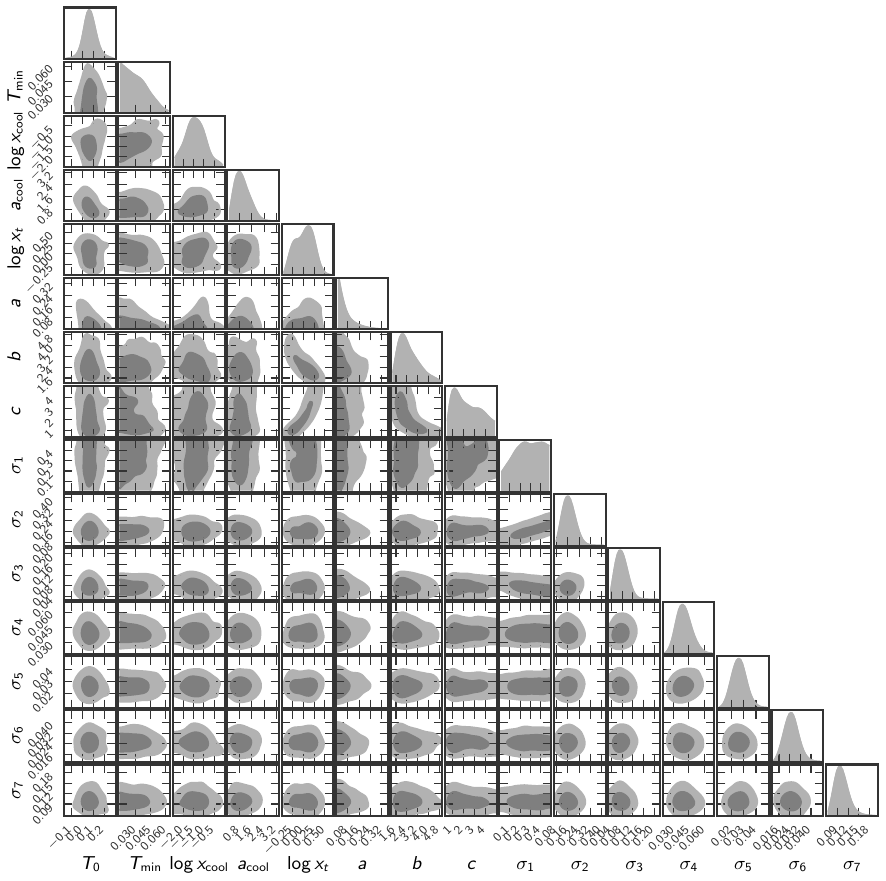}
\caption{\footnotesize Marginalised posterior likelihood for the parameters of the best-fitting temperature profile model to the \excpres\ data.  }
\label{f:corner}
\end{figure*}

\end{appendix}
\end{document}